\documentclass[aps,PhysRevB,reprint,superscriptaddress]{revtex4-2}

\usepackage[T1]{fontenc}
\usepackage{mathptmx}

\usepackage{graphicx}

\usepackage[version=4]{mhchem}
\usepackage{textcomp}
\usepackage{dcolumn}
\usepackage{bm}
\usepackage{esdiff}
\usepackage{natbib}
\usepackage[colorlinks=true,citecolor=blue,linkcolor=blue,urlcolor=blue,pdfhighlight=/O]{hyperref}
\begin{document}
\title{Epitaxial stabilization of an orthorhombic Mg-Ti-O superconductor}

\author{Zhuang Ni}
\affiliation{Beijing National Laboratory for Condensed Matter Physics, Institute of Physics, Chinese Academy of Sciences, Beijing 100190, China}
\affiliation{School of Physical Sciences, University of Chinese Academy of Sciences, Beijing 100049, China}

\author{Wei Hu}
\author{Qinghua Zhang}
\author{Yanmin Zhang}
\author{Peiyu Xiong}
\author{Qian Li}\email{qianli@iphy.ac.cn}
\affiliation{Beijing National Laboratory for Condensed Matter Physics, Institute of Physics, Chinese Academy of Sciences, Beijing 100190, China}

\author{Jie Yuan}
\affiliation{Beijing National Laboratory for Condensed Matter Physics, Institute of Physics, Chinese Academy of Sciences, Beijing 100190, China}
\affiliation{Songshan Lake Materials Laboratory, Dongguan, Guangdong 523808, China}

\author{Qihong Chen}
\author{Beiyi Zhu}
\author{Hua Zhang}
\affiliation{Beijing National Laboratory for Condensed Matter Physics, Institute of Physics, Chinese Academy of Sciences, Beijing 100190, China}

\author{Xiaoli Dong}
\author{Lin Gu}
\author{Kui Jin}
\affiliation{Beijing National Laboratory for Condensed Matter Physics, Institute of Physics, Chinese Academy of Sciences, Beijing 100190, China}
\affiliation{School of Physical Sciences, University of Chinese Academy of Sciences, Beijing 100049, China}
\affiliation{Songshan Lake Materials Laboratory, Dongguan, Guangdong 523808, China}

\begin{abstract}
The family of titanium oxide superconductors exhibits many intriguing phenomena comparable to cuprates and iron pnictides/chalcogenides, and thus provides an ideal platform to contrastively study the unconventional pairing mechanism of high-temperature superconductors. Here, we successfully deposit superconducting Mg-Ti-O films on \ce{MgAl2O4} substrates with three principal orientations by ablating a \ce{MgTi2O4} target. Particularly, it is striking to observed that a single-crystalline film of an unintended structure has been grown on the (011)-oriented substrate, with the highest zero resistance transition temperature ($T_{\mathrm{c}0}$) of 5.0 K among them. The film has a highly reduced Mg/Ti ratio and an orthorhombic \ce{Ti9O10}-like structure (denoted as Mg: \ce{Ti9O10}), demonstrated by further characterizations of chemical composition and structure. Such a structure is unstable in bulk but favorable to be epitaxially stabilized on the (011)-surface of \ce{MgAl2O4} due to a relatively small strain at the formed interface. An isotropic upper critical field ($B_{\mathrm{c}2}$) up to 13.7 T that breaks the Pauli limit is observed in the Mg: \ce{Ti9O10} film, analogous to other superconducting titanium oxides. The similarity points to a common origin for the superconductivity in the family, which will provide valuable opinions for the mechanism of unconventional superconductivity in transition metal compounds. 
\end{abstract}

\maketitle

The strong coupling among charge, spin, orbital, and lattice of $3d$ electrons in transition metal compounds always gives rise to various emergent phenomena \cite{RevModPhys.77.871,doi:10.1126/science.1107559,RevModPhys.83.471,RN109,Pines_2016}. One typical example is the high-temperature superconductivity discovered in cuprates \cite{RN110} and iron pnictides/chalcogenides \cite{doi:10.1021/ja800073m}. It seems confusing to observe the coexistence of superconductivity and antiferromagnetic ordering in both families \cite{RN116,PhysRevB.78.184512,RN125}. However, recent investigations indicate that the antiferromagnetic interaction among $3d$ electrons is likely to be the origin of high-temperature superconductivity \cite{Anderson_2004,PhysRevLett.101.206404,RevModPhys.84.1383,RN111,RevModPhys.85.849}. Due to the complexity of these systems, it is still puzzling why high-temperature superconductivity can be achieved in the specific families of $3d$ transition metal compounds. Thus, study on a reference system, of which superconductivity also originates from $3d$ electrons, will provide valuable perspectives for further comprehension of unconventional superconductivity.

The family of titanium oxide superconductors (TOS) is an ideal candidate for contrastive study of high-temperature superconductivity. Many intriguing phenomena similar to that of high-temperature superconductors have been unveiled in TOS, such as pseudogap \cite{RN28} and exotic superconductor-metal \cite{RN114} or -insulator \cite{PhysRevB.98.064501,RN34,PhysRevB.103.L140501} transition. Although TOS have various structures, e.g., cubic spinel \ce{LiTi2O4} \cite{JOHNSTON1973777}, cubic rock-salt TiO \cite{doi:10.1021/acsomega.7b00048}, triclinic \ce{Ti4O7} \cite{RN112}, and monoclinic $\gamma$-\ce{Ti3O5} \cite{RN112}, they have a notably common structural unit, Ti-O bond. Similar to the \ce{CuO2} layer in cuprates or FeAs/Se layer in iron-based superconductors, Ti-O bond seems also crucial to the superconductivity of titanium oxides. In the case of \ce{LiTi2O4}, the superconductor with the highest superconducting transition temperature $T_{\mathrm{c}}\sim 13$ K among the family, the substantial Ti-O hybridization \cite{PhysRevB.38.11352} contributes remarkably to the superconductivity \cite{Chen_2011}. In contrast, stoichiometric TiO exhibits a much lower $T_{\mathrm{c}}\sim 0.5$ K than oxygen-rich \ce{TiO_${1+\delta}$} samples (up to $\sim 7$ K) \cite{doi:10.1021/acsomega.7b00048,zhang2017enhanced,RN114,PhysRevB.98.064501,FAN2019607} due to the direct Ti-Ti bonding \cite{doi:10.1126/sciadv.abd4248}. Besides, the coexistence of superconductivity with other collective excitations, e.g., orbital-related state in \ce{LiTi2O4} \cite{RN3} and ferromagnetism in Mg-doped TiO \cite{FAN202066}, also indicates that the family provides a promising window for superconductivity in the vicinity of a competing order, other than cuprates \cite{RN116} and iron-based superconductors \cite{RN115}. Particularly, superconducting transition up to $\sim 5$ K can be achieved by suppressing the orbital ordering in \ce{MgTi2O4} \cite{PhysRevB.101.220510}, once known to be a band insulator with a helical dimerization pattern of alternating short and long Ti-Ti bonds \cite{PhysRevLett.92.056402,PhysRevLett.93.077208,PhysRevLett.94.156402,PhysRevB.78.125105}. Therefore, further study on TOS is desired to thoroughly understand the mechanism of superconductivity originated from $3d$ electrons.

Nevertheless, extensive study on TOS is hampered by the lack of high-qualified single crystal sample owning to the thermodynamic or chemical instability of their crystal lattice \cite{doi:10.1126/sciadv.abd4248,10.1111/j.1151-2916.1999.tb02245.x}. Fortunately, many metastable or unstable phases can be stabilized in the form of single-crystalline films by the means of epitaxial stabilization \cite{doi:10.1021/cm021111v,RN122,Ramesh2019NRM,RN123,RN124}. Exotic behaviors comparable to high-temperature superconductors have been exhibited in the films of TOS \cite{RN28,RN114,PhysRevB.98.064501,RN34,PhysRevB.103.L140501,RN3,FAN202066}. Moreover, some new TOS, e.g., \ce{Ti4O7} \cite{RN112}, $\gamma$-\ce{Ti3O5} \cite{RN112}, orthorhombic \ce{Ti2O3} \cite{RN118}, and \ce{MgTi2O4} \cite{PhysRevB.101.220510}, which have never been reported to be superconducting in bulk, are disclosed by thin film deposition. Besides providing an elastic strain, the substrate can also tailor the properties of the film much by its crystallographic direction \cite{PhysRevApplied.1.051002,PhysRevB.95.054510,PhysRevLett.127.086804}. Due to the strong-coupling nature of TOS, the crystal and/or electronic structure of the film may be sensitive to the crystallographic symmetry of the substrate surface. Therefore, it is necessary to study their films grown on substrates with different orientations, where the emergence of different structural or electronic phases is promising.

In this Letter, Mg-Ti-O films with complete superconducting transitions are deposited by ablating a stoichiometric \ce{MgTi2O4} target on \ce{MgAl2O4} (MAO) substrates with three principal orientations. However, the phases formed in the obtained films exhibit a remarkable orientational selectivity. In contrast to the spinel \ce{MgTi2O4} (space group: $Fd\overline{3}m$) phase on (001)-oriented MAO, a highly Mg-deficient phase with a higher $T_{\mathrm{c}}$ but in a strikingly distinct structure emerges on the substrate with (011) orientation. Thorough structural characterizations demonstrate the orthorhombic \ce{Ti9O10}-like structure of the film. Such a structure has a smaller mismatch with the (011)-oriented MAO substate than spinel \ce{MgTi2O4}, so that its formation is unintended but reasonable. 

The Mg-Ti-O superconducting films are deposited under high vacuum by pulsed laser deposition (PLD) using a KrF excimer laser ($\lambda=248$ nm) and a commercial stoichiometric \ce{MgTi2O4} target. The chamber is evacuated to a base pressure better than $1\times 10^{-6}$ Torr before growth. During deposition, the laser pulse energy, repetition rate, and grown temperature are fixed at 400 mJ, 4 Hz, and 840\textcelsius, respectively. Films are deposited on the MAO substrates with (001), (111), and (011) orientations in the same batch to ensure the uniformity of some crucial growth parameters. The thicknesses of the samples used for subsequent characterizations are $\sim 150$ nm. Structural characterizations of the films are conducted by an x-ray diffractometer (XRD) at room temperature using Cu $K_{\alpha 1}$ radiation ($\lambda=1.54056$ {\AA}). The electrical and magnetic transport properties are measured using Physical Property Measurement System (PPMS). The magnetization measurements are performed by the Magnetic Property Measurement System (MPMS). The high-angle annular dark-field scanning transmission electron microscopy (HAADF-STEM) images are collected to probe the microstructures for cross-sectional samples. 

\begin{figure}[htbp]
\includegraphics[width=\linewidth]{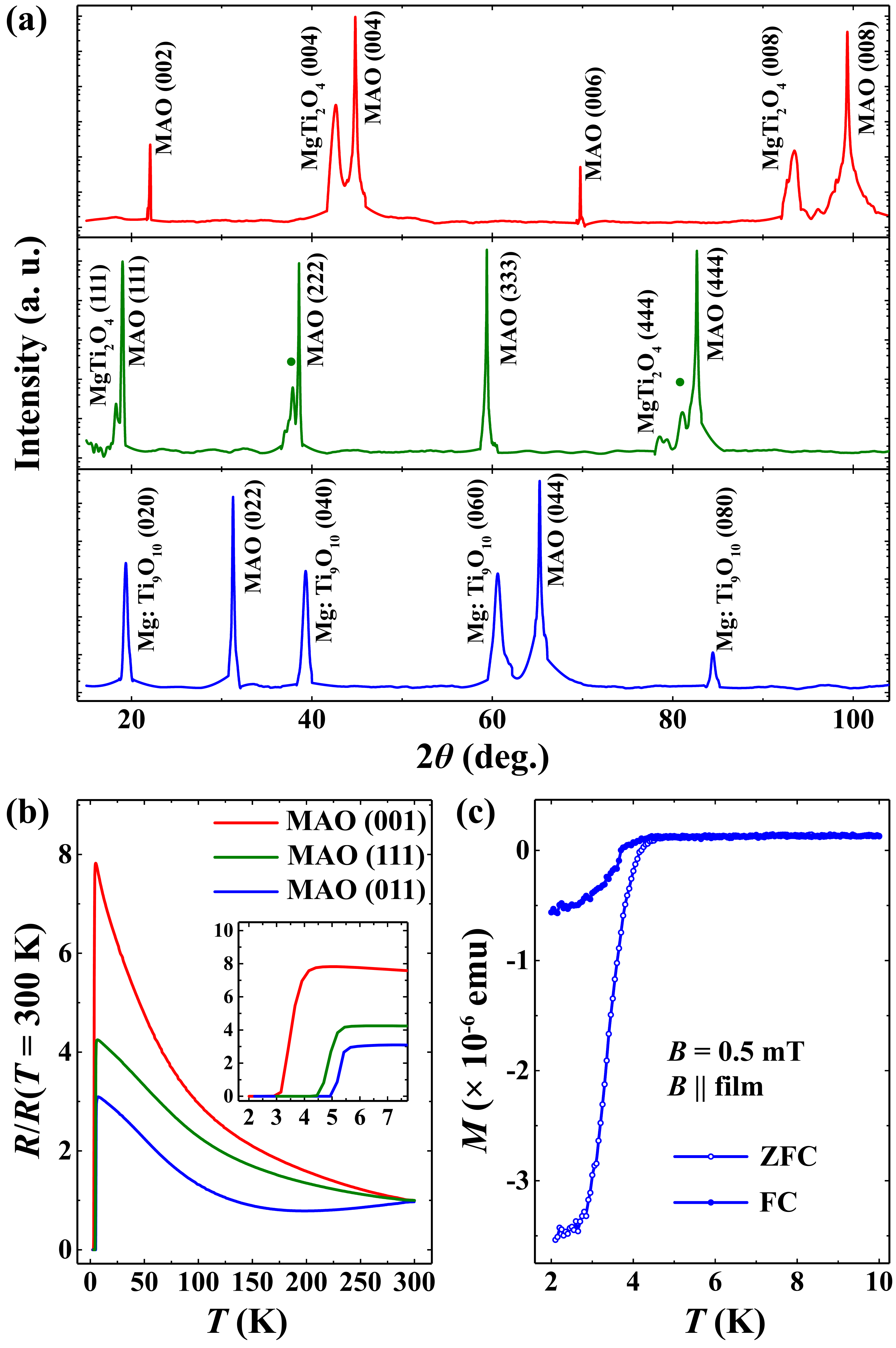}
\caption{\label{Fig1} (a) Out-of-plane XRD spectra of $\theta$-$2\theta$ scanning for Mg-Ti-O films on (001)- (top panel), (111)- (middle panel), and (011)-oriented (bottom panel) MAO substrates. All the diffraction peaks belonging to the films grown on (001)- and (011)-oriented MAO substrates can be indexed by spinel \ce{MgTi2O4} and orthorhombic Mg: \ce{Ti9O10} structure, respectively, while unidentified phases except for spinel \ce{MgTi2O4} (denoted by solid spheres and squares) form on (111)-oriented MAO substrate. (b) Temperature dependence of normalized electrical resistance for films grown on MAO substrates with three different orientations. Inset: Enlarged low-temperature resistance around the superconducting transitions. (c) Temperature dependence of magnetization of the Mg-Ti-O film deposited on MAO (011) substrate at 0.5 mT with and without field cooling. }
\end{figure}

As shown in Fig. \ref{Fig1}(b), complete superconducting transitions can be seen in all the $R(T)$ curves of Mg-Ti-O films deposited on (001)-, (111)-, and (011)-oriented MAO substrates, with $T_{\mathrm{c}0}$'s of 2.6, 4.2, and 5.0 K, respectively. Nevertheless, the results of structural characterization exhibited in Fig. \ref{Fig1}(a) suggest they are in different structural phases. In the top panel, out-of-plane reflections corresponding to the \ce{MgTi2O4} ($00l$) orientation only can be observed beside the peaks from the substrate, suggesting that ($00l$)-oriented spinel \ce{MgTi2O4} film without impurity phase has been deposited on MAO (001) substrate. It should be noted that superconducting \ce{MgTi2O4} film with $T_{\mathrm{c}0}$ can be achieved in a single-layered film, rather than in an engineered superlattice geometry \cite{PhysRevB.101.220510}, by further optimization of deposition process. For the film deposited on (111)-oriented MAO substrate [middle panel of Fig. \ref{Fig1}(a)], diffraction peaks differ from those of (111)-type \ce{MgTi2O4} and MAO are shown, indicating the formation of an eutectic film containing spinel \ce{MgTi2O4} and other phase(s). Dramatically, sharp diffraction peaks from the film on MAO (011) substrate located at $2\theta=19.37^{\circ}$, $39.30^{\circ}$, $60.58^{\circ}$, and $84.58^{\circ}$, can be observed in the XRD pattern shown in the bottom panel of Fig. \ref{Fig1}(a). The locations of these peaks are significantly distinct from those of spinel \ce{MgTi2O4} phase. However, the sines of the diffraction angles conform to a ratio of $1:2:3:4$, suggesting that they stem from a single-crystalline phase according to Bragg's law. Superconductivity of the film is also evidenced by magnetization measurements. The temperature dependence of the magnetic susceptibility in both zero-field-cooling (ZFC) and field-cooling (FC) modes discloses that the Meissner state appears at $\sim 4.5$ K, as seen in Fig. \ref{Fig1}(c), consistent with the $T_{\mathrm{c}0}$ obtained by transport measurements. In other words, a single-phased superconducting film different from \ce{MgTi2O4} but with a higher $T_{\mathrm{c}}$ has been obtained on MAO (011) substrate, albeit a \ce{MgTi2O4} target is employed for ablating. It is surprising that superconductivity can be achieved in a different phase via the orientational tuning effect of the substrate, and therefore further investigation on composition and structure is warranted.

In order to exclude the possible affection from the substrate in composition characterization, Mg- and Ti-free \ce{(La,Sr)(Al,Ta)O3} (LSAT) substrates with different crystallographic directions are also used to deposit Mg-Ti-O films. The structural and transport properties of the films grown on LSAT substrates are quite analogous to those of the films grown on MAO substrates [see Supplemental Material \cite{supplemental}, Figs. S1(a) and S1(b)]. Therefore, the analysis of composition for the samples grown on LSAT substrates can be regarded as a reference. It is strikingly observed that the film grown on LSAT (011) substrate has an enhanced Ti/Mg ratio of $\times 3.75$ compared with the target [see Supplemental Material \cite{supplemental}, Fig. S1(c)]. Such a remarkable deviation between the target and the film stoichiometry seems conflicting with the well-known stoichiometric transfer feature of PLD. Actually, incongruent ablation can be induced in many systems, e.g., \ce{SrTiO3} \cite{doi:10.1063/1.4754112}, \ce{LaAlO3} \cite{PhysRevLett.110.196804}, and \ce{La_{0.4}Ca_{0.6}MnO3} \cite{10.1002/admi.201701062},  by changing the laser fluence, mainly due to the difference in cohesive energy and the atomic mass among different elements \cite{PhysRevMaterials.1.073402,SCHOU20095191}. Meanwhile, preferential scattering of the lighter atoms by the background gas often induces the relative enrichment of heavier elements in films \cite{10.1002/admi.201701062,PhysRevLett.111.036101}. Moreover, many other factors, e.g., possible chemical interaction between the plasma plume and the background gas \cite{10.1002/admi.201701062}, elemental transfer from the substrate to the film \cite{doi:10.1063/1.3515849}, and the resputtering or backscattering \cite{10.1002/admi.201701062,SCHOU20095191}, can also influence the composition of film. Nevertheless, these mechanisms cannot point to such a dramatic loss of Mg content in the films grown on (011)-oriented substrates, indicating an essentially different scenario.

The Raman and in-plane XRD spectra are collected to clarify the structure of the film on MAO (011) substrate. The Raman spectrum of the film exhibits similar feature as the superconducting orthorhombic \ce{Ti2O3} \cite{RN118}, implying that the films grown on MAO (011) substrate should also have an orthorhombic structure (see Supplemental Materials \cite{supplemental} for more details). However, a significant deviation exists between the locations of diffraction peaks predicted by the lattice constant of orthorhombic \ce{Ti2O3} \cite{RN118} and the actual values in Fig. \ref{Fig1}(a), which cannot be interpreted as strain effect induced by the substrate. By carefully searching for titanium oxides in the Inorganic Crystal Structure Database (ICSD) following the structural information above, the \ce{Ti9O10} compound (ICSD 77698) enters our consciousness. The \ce{Ti9O10} phase is proposed by Hilti \cite{RN120}, with an orthorhombic structure (space group: $Immm$) and lattice constants of $a=3.986$ {\AA}, $b=9.086$ {\AA}, and $c=2.981$ {\AA}. Accordingly, the locations of diffraction peaks corresponding to the (020), (040), (060), and (080) planes of \ce{Ti9O10} are estimated to be $19.53^{\circ}$, $39.64^{\circ}$, $61.14^{\circ}$, and $85.41^{\circ}$, respectively, fairly close to the values of the film deposited on MAO (011) substrate in Fig. \ref{Fig1}(a). Meanwhile, the (200) and ($0\overline{2}2$) diffractions of \ce{Ti9O10} should be appeared at $45.47^{\circ}$ and $62.23^{\circ}$, respectively, also in highly agreement with the in-plane XRD results (see Supplemental Material \cite{supplemental}, Fig. S3). Therefore, it is reasonable to speculate that the superconducting Mg-Ti-O film deposited on (011)-oriented MAO substrate are isostructural to \ce{Ti9O10}.

\begin{figure}[htbp]
\includegraphics[width=\linewidth]{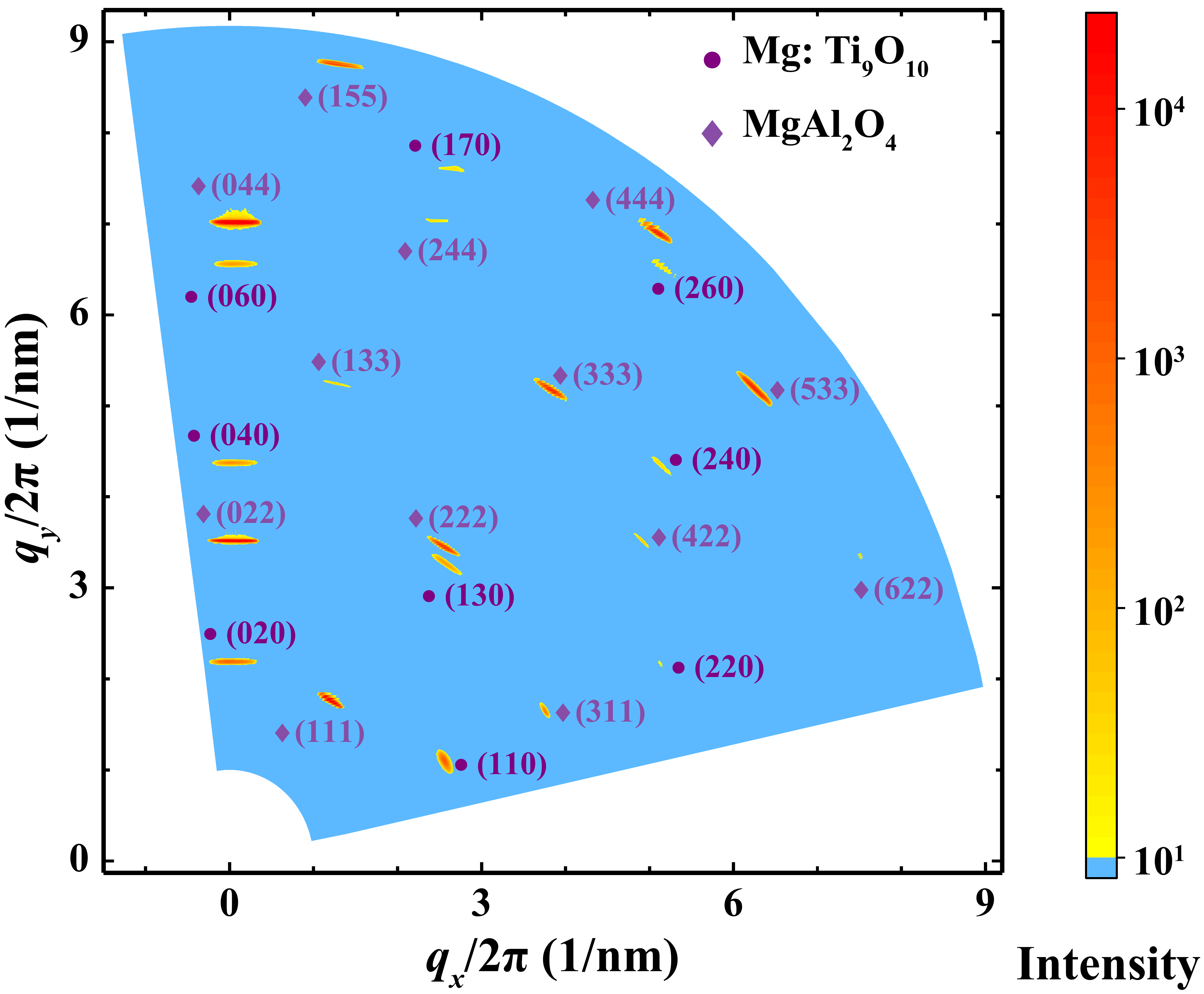}
\caption{\label{Fig2} The wide range reciprocal space mapping of the Mg-Ti-O film grown on MAO (011) substrate. The recorded intensity is plotted in the reciprocal space coordinates, where the horizontal axis $q_x$ and vertical axis $q_y$ correspond to the $[100]$ and $[010]$ directions of the film, respectively. The diamonds and solid spheres denote the planes belonging to the MAO substrate and Mg: \ce{Ti9O10} film, respectively.}
\end{figure}

Although the crystal structure has been predicted for half a century \cite{RN120}, stable \ce{Ti9O10} bulk with high-purity is difficult to be synthesized for the inevitable coexistence with other titanium oxides, e.g., \ce{Ti3O5} and \ce{TiO2} (anatase) \cite{Valeeva2017NPCM,RN119}. Nevertheless, the structure and phase purity of the film can be further confirmed by the wide-range reciprocal space mappings (RSMs) \cite{Inaba2013AMPC}. A series of $2\theta$-$\omega$ scans are performed along with the step-tilting of the $\chi$ axis which lies parallel to the $[001]$ axis of the film. Theoretically, all the ($hk0$) planes of the film (indexed by the \ce{Ti9O10}-like structure), along with the \ce{MgAl2O4} ($HKK$) planes, within the measurement range should be detected. The intensity of the wide-range RSM is recorded by a 2D detector and displayed in the reciprocal space coordinates (Fig. \ref{Fig2}). All the diffraction peaks in the wide-range RSM can be indexed by either the \ce{Ti9O10}-like structure or the MAO substrate, without any signal of impurity phase, suggesting that a pure orthorhombic \ce{Ti9O10}-like phase is obtained with the assistance of strain from the substrate. Additionally, the lattice constants along the $[100]$ and $[010]$ directions of the film are calculated to be $a=3.925$ {\AA} and $b=9.140$ {\AA} using the wide-range RSM results, respectively. These values are acquired with more diffraction peaks included, and thus we take them as the lattice constants of the film in the following discussion (along with $c=2.995$ {\AA} obtained by in-plane XRD, see Supplemental Material \cite{supplemental}). 

\begin{figure}[htbp]
\includegraphics[width=\linewidth]{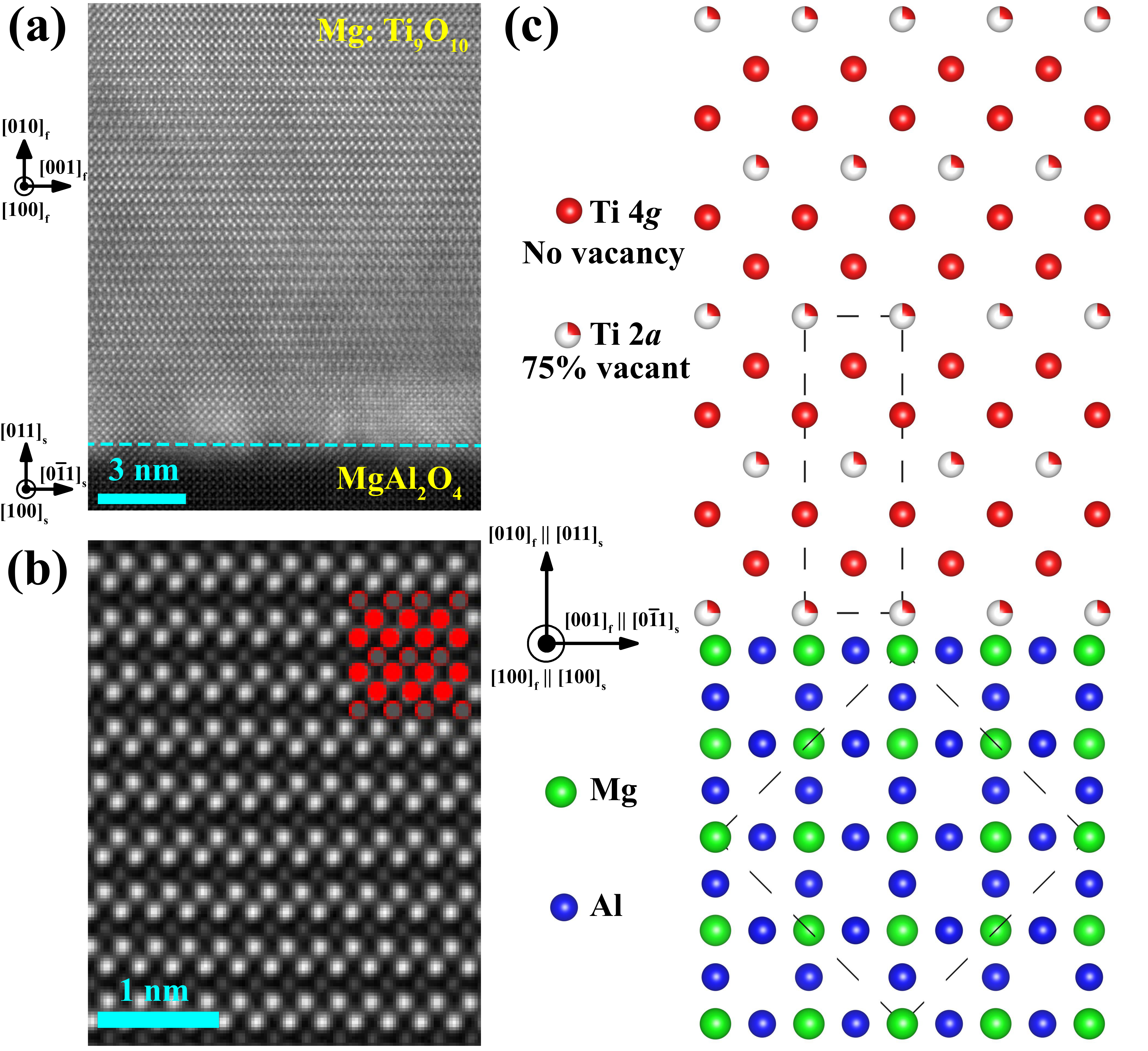}
\caption{\label{Fig3} Microstructure of Mg-Ti-O film grown on MAO (011) substrate. (a) HAADF-STEM image captured around the interface (denoted by blue dash line) between film and substrate. (b) Zoom-in HAADF-STEM image of the film region. The bright spots and dark spots at the top right-hand corner are labeled by red and grey spheres, respectively. (c) Illustration of the epitaxial relationship for Mg: \ce{Ti9O10}/MAO (011) sample. The framework of Ti sublattice in an ideal orthorhombic \ce{Ti9O10} structure are exhibited in the top part, where the existence of few Mg atoms is not included. The red-white spheres represent the Ti atoms located at $2a$ positions which are filled with vacancies by 75\%, and red atoms denote the Ti atoms at $4g$ positions which are vacancy-free. }
\end{figure}

A significant feature of orthorhombic \ce{Ti9O10} structure, i.e., vacancy ordering, can also be observed in the film grown on MAO (011) substrate through HAADF-STEM image, which further confirms our speculation. The overlayer framework of Ti atoms on the (100) plane of \ce{Ti9O10} are presented in the top part of Fig. \ref{Fig3}(c), where two vacancy-free planes composed of Ti $4g$ and O $4h$ positions are alternate with one vacancy plane formed by Ti $2a$ and O $2c$ positions (filled with vacancies by 75\% and 50\%, respectively) \cite{RN120,Valeeva2017NPCM,RN119}. The film and substrate regions are clearly recognizable in the HAADF-STEM image [Fig. \ref{Fig3}(a)], inserted by a $\sim 4$ nm-thick transition layer. Figure \ref{Fig3}(b) exhibits the zoom-in HAADF-STEM image of the film region, in which two distinct types of spots, divided by their brightness contrast, can be observed. It is known that the brightness of each spot is positively correlated to the atomic number ($Z$) and/or the occupation rate of the corresponding atomic column. Considering the much larger $Z=22$ of Ti than Mg ($Z=12$) and O ($Z=8$), as well as the highly reduced Mg/Ti ratio in the film, the spots in the HAADF-STEM image are believed to be the reflection of Ti sublattice. The arrangement of the spots presents an obvious periodicity, in which two lines filled with bright spots are alternate with one line composed of dark spots. Such a periodicity is in well consistent with the framework of Ti sublattice in \ce{Ti9O10} structure, as labeled in the top right-hand corner of Fig. \ref{Fig3}(b). As a reference, the lattice constants along $[001]$ and $[010]$ directions are estimated to be 2.901 {\AA} and 9.121 {\AA} by calculating the average distances of corresponding spots, respectively. The deviation between the values collected from XRD and STEM results is reasonable so that the \ce{Ti9O10}-like structure of the film can be verified again. Although it is still difficult to determine the role played by the few Mg atoms, i.e., substitutional or interstitial, the phase of the film can be affirmatively denoted as Mg: \ce{Ti9O10}. As illustrated in Fig. \ref{Fig3}(c), the epitaxial relationship between the film and the MAO (011) substrate is determined to be $[100]$ Mg: \ce{Ti9O10}$\parallel [100]$ MAO and $[001]$ Mg: \ce{Ti9O10}$\parallel [0\overline{1}1]$ MAO, with lattice mismatches of $-2.96\%$ and $+4.62\%$, respectively. Such mismatches are much smaller than that between \ce{MgTi2O4} and MAO substrate, seeming to be responsible for the selective phase formation in the film. We have also discussed the Mg-Ti-O phases formed in the films deposited on other substrates (see Supplemental Material \cite{supplemental} for details). It should be noted that less-strained interfaces are favorable in the obtained samples. During PLD process, the atoms ablated from the target have enough kinetic energies to rearrange themselves into a better matched phase, aiming to form a coherent interface with a reduced strain energy \cite{doi:10.1021/cm021111v,doi:10.1021/acsami.6b01630}. Therefore, it is preferential to form Mg: \ce{Ti9O10} phase rather than spinel \ce{MgTi2O4} at the (011)-surface of MAO substrate. 

\begin{figure}[htbp]
\includegraphics[width=\linewidth]{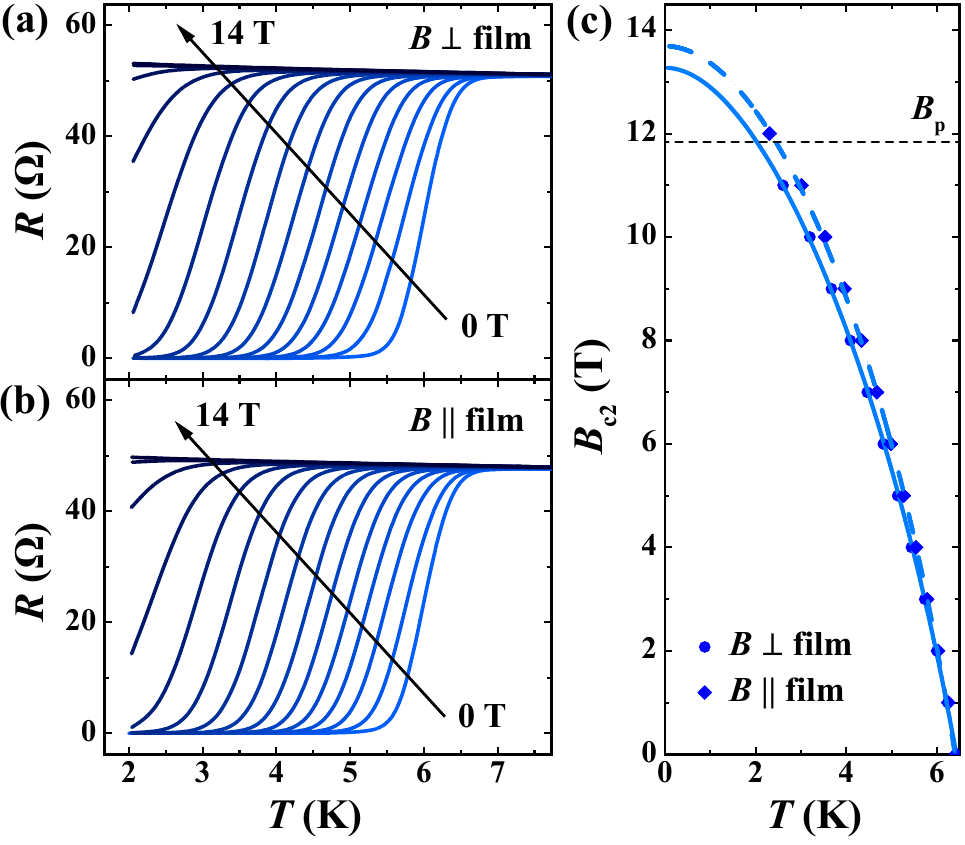}
\caption{\label{Fig4} (a), (b) Temperature dependence of the electrical resistance of the Mg: \ce{Ti9O10}/MAO (011) sample under various magnetic fields perpendicular (a) and parallel (b) to the film; (c) Temperature-dependent $B_{\mathrm{c}2}$ of the Mg: \ce{Ti9O10}/MAO (011) sample when $B\perp$ film (solid squares) and $B\parallel$ film (solid circle). Solid and dashed lines are fitted by the WHH theory for $B\perp$ film and $B\parallel$ film, respectively. The Pauli limit $B_{\mathrm{p}}$ predicted by the weak-coupling BCS paramagnetic formula is labeled by the dashed horizontal line. }
\end{figure}

The characteristic superconducting features of the Mg: \ce{Ti9O10} film are also examined by magnetotransport measurements. The temperature dependent electrical resistance is measured under a series of magnetic fields, from 0 to 14 T. As shown in Fig. \ref{Fig4}(a), the magnetic fields applied perpendicular to the film ($B\perp$ film) gradually suppress the superconductivity and the superconducting transition is pushed to lower temperatures. Almost identical magnetoresistance behavior is observed while $B\parallel$ film [Fig. \ref{Fig4}(b)], indicating an isotropic $B_{\mathrm{c}2}$. The values of $B_{\mathrm{c}2}(T)$, determined using a criterion of 90\% of the normal state resistance, can be well fitted by the Werthamer-Helfand-Hohenberg (WHH) theory \cite{PhysRev.147.295} if spin paramagnetism and the spin-orbit interaction are taken into consideration, as seen in Fig. \ref{Fig4}(c). The zero temperature upper critical field $B_{\mathrm{c}2}(T=0)$ is calculated to be 13.3 T and 13.7 T in the case of $B\perp$ film and $B\parallel$ film, respectively, strikingly breaking the Pauli limit $B_{\mathrm{p}}\sim 12$ T predicted by the weak-coupling BCS paramagnetic formula $B_{\mathrm{p}}=1.84T_{\mathrm{c}}$ ($T_{\mathrm{c}}$ is taken as the temperature at which the resistance crosses 90\% of the normal state resistance). Such a relatively high $B_{\mathrm{c}2}$ of isotropy mimics the behaviors of some other TOS \cite{zhang2017enhanced,PhysRevB.101.220510,PhysRevB.100.184509}, which possibly stems from the special electronic structures correlated to the strong coupling of $3d$ electrons \cite{PhysRevB.100.184509}. Additionally, the Ginzburg-Landau coherence length $\xi_{\mathrm{GL}}$ is estimated to be $\sim 4.01$ nm following WHH formula $\xi_{\mathrm{GL}}=\sqrt{\phi_0/(2\pi B_{\mathrm{c}2}^{\mathrm{orb}})}$, where $B_{\mathrm{c}2}^{\mathrm{orb}}=-0.69T_{\mathrm{c}}(dB_{\mathrm{c}2}/dT)|_{T=T_{\mathrm{c}}}$ is the orbital limited upper critical field \cite{PhysRev.147.295}.

Overall, Mg-Ti-O films with complete superconducting transitions are successfully deposited on MAO substrates with different orientations using a \ce{MgTi2O4} target. Notably, a single-crystalline film with a structure distinct from spinel \ce{MgTi2O4} is formed on MAO (011), presenting the highest $T_{\mathrm{c}0}$ of 5.0 K among these samples. Thorough characterizations reveals that the film is in an orthorhombic \ce{Ti9O10}-like structure with a highly reduced Mg content. Although such a structure is unstable in bulk, it can be epitaxially stabilized on the (011)-surface of MAO due to the relatively small strain at the obtained interface. Our work not only introduces a different member to the TOS family, but also demonstrates the considerable potenial of film deposition in exploring more superconductors or other functional materials via epitaxial stabilization. Furthermore, an isotropic $B_{\mathrm{c}2}$ breaking the Pauli limit is exhibited in the Mg: \ce{Ti9O10} film, resembling those of some other TOS \cite{zhang2017enhanced,PhysRevB.101.220510,PhysRevB.100.184509}. Considering their common structural unit, Ti-O bond, it should be believed that some ingredients not relying on specific structure may act as the superconducting \emph{gene} \cite{HU2016561} in the family. Cracking such a gene will bring us different opinion for the unconventional superconductivity originated from $3d$ electrons, and therefore the discovery of TOS with higher $T_{\mathrm{c}}$ is promising.

\begin{acknowledgements}
We thank Lihong Yang for assistance in experimental measurements. This work is supported by the National Key R\&D Program of China (Grants No. 2017YFA0302902, No. 2017YFA0303003, No. 2018YFB0704102, and No. 2021YFA0718700), the National Natural Science Foundation of China (Grants No. 11804378, No. 11834016, No. 11961141008, No. 11927808, No.52072400, No. 12174428, and No. 12104490), the Strategic Priority Research Program (B) of Chinese Academy of Sciences (Grants No. XDB25000000, and No. XDB33000000), the Beijing Natural Science Foundation (Grants No. Z190008, and No. Z190010), CAS Interdisciplinary Innovation Team, Key-Area Research and Development Program of Guangdong Province (Grant No. 2020B0101340002), and the Center for Materials Genome. Y.Z. thanks the support from the China Postdoctoral Science Foundation (Grant No. 2020M680729).
\end{acknowledgements}

\bibliography{Mg-Ti-O-011} 

\begin{thebibliography}{64}%
\makeatletter
\providecommand \@ifxundefined [1]{%
 \@ifx{#1\undefined}
}%
\providecommand \@ifnum [1]{%
 \ifnum #1\expandafter \@firstoftwo
 \else \expandafter \@secondoftwo
 \fi
}%
\providecommand \@ifx [1]{%
 \ifx #1\expandafter \@firstoftwo
 \else \expandafter \@secondoftwo
 \fi
}%
\providecommand \natexlab [1]{#1}%
\providecommand \enquote  [1]{``#1''}%
\providecommand \bibnamefont  [1]{#1}%
\providecommand \bibfnamefont [1]{#1}%
\providecommand \citenamefont [1]{#1}%
\providecommand \href@noop [0]{\@secondoftwo}%
\providecommand \href [0]{\begingroup \@sanitize@url \@href}%
\providecommand \@href[1]{\@@startlink{#1}\@@href}%
\providecommand \@@href[1]{\endgroup#1\@@endlink}%
\providecommand \@sanitize@url [0]{\catcode `\\12\catcode `\$12\catcode
  `\&12\catcode `\#12\catcode `\^12\catcode `\_12\catcode `\%12\relax}%
\providecommand \@@startlink[1]{}%
\providecommand \@@endlink[0]{}%
\providecommand \url  [0]{\begingroup\@sanitize@url \@url }%
\providecommand \@url [1]{\endgroup\@href {#1}{\urlprefix }}%
\providecommand \urlprefix  [0]{URL }%
\providecommand \Eprint [0]{\href }%
\providecommand \doibase [0]{https://doi.org/}%
\providecommand \selectlanguage [0]{\@gobble}%
\providecommand \bibinfo  [0]{\@secondoftwo}%
\providecommand \bibfield  [0]{\@secondoftwo}%
\providecommand \translation [1]{[#1]}%
\providecommand \BibitemOpen [0]{}%
\providecommand \bibitemStop [0]{}%
\providecommand \bibitemNoStop [0]{.\EOS\space}%
\providecommand \EOS [0]{\spacefactor3000\relax}%
\providecommand \BibitemShut  [1]{\csname bibitem#1\endcsname}%
\let\auto@bib@innerbib\@empty
\bibitem [{\citenamefont {Levin}\ and\ \citenamefont
  {Wen}(2005)}]{RevModPhys.77.871}%
  \BibitemOpen
  \bibfield  {author} {\bibinfo {author} {\bibfnamefont {M.}~\bibnamefont
  {Levin}}\ and\ \bibinfo {author} {\bibfnamefont {X.-G.}\ \bibnamefont
  {Wen}},\ }\href {https://doi.org/10.1103/RevModPhys.77.871} {\bibfield
  {journal} {\bibinfo  {journal} {Rev. Mod. Phys.}\ }\textbf {\bibinfo {volume}
  {77}},\ \bibinfo {pages} {871} (\bibinfo {year} {2005})}\BibitemShut
  {NoStop}%
\bibitem [{\citenamefont {Dagotto}(2005)}]{doi:10.1126/science.1107559}%
  \BibitemOpen
  \bibfield  {author} {\bibinfo {author} {\bibfnamefont {E.}~\bibnamefont
  {Dagotto}},\ }\href {https://doi.org/10.1126/science.1107559} {\bibfield
  {journal} {\bibinfo  {journal} {Science}\ }\textbf {\bibinfo {volume}
  {309}},\ \bibinfo {pages} {257} (\bibinfo {year} {2005})}\BibitemShut
  {NoStop}%
\bibitem [{\citenamefont {Basov}\ \emph {et~al.}(2011)\citenamefont {Basov},
  \citenamefont {Averitt}, \citenamefont {van~der Marel}, \citenamefont
  {Dressel},\ and\ \citenamefont {Haule}}]{RevModPhys.83.471}%
  \BibitemOpen
  \bibfield  {author} {\bibinfo {author} {\bibfnamefont {D.~N.}\ \bibnamefont
  {Basov}}, \bibinfo {author} {\bibfnamefont {R.~D.}\ \bibnamefont {Averitt}},
  \bibinfo {author} {\bibfnamefont {D.}~\bibnamefont {van~der Marel}}, \bibinfo
  {author} {\bibfnamefont {M.}~\bibnamefont {Dressel}},\ and\ \bibinfo {author}
  {\bibfnamefont {K.}~\bibnamefont {Haule}},\ }\href
  {https://doi.org/10.1103/RevModPhys.83.471} {\bibfield  {journal} {\bibinfo
  {journal} {Rev. Mod. Phys.}\ }\textbf {\bibinfo {volume} {83}},\ \bibinfo
  {pages} {471} (\bibinfo {year} {2011})}\BibitemShut {NoStop}%
\bibitem [{\citenamefont {Hwang}\ \emph {et~al.}(2012)\citenamefont {Hwang},
  \citenamefont {Iwasa}, \citenamefont {Kawasaki}, \citenamefont {Keimer},
  \citenamefont {Nagaosa},\ and\ \citenamefont {Tokura}}]{RN109}%
  \BibitemOpen
  \bibfield  {author} {\bibinfo {author} {\bibfnamefont {H.~Y.}\ \bibnamefont
  {Hwang}}, \bibinfo {author} {\bibfnamefont {Y.}~\bibnamefont {Iwasa}},
  \bibinfo {author} {\bibfnamefont {M.}~\bibnamefont {Kawasaki}}, \bibinfo
  {author} {\bibfnamefont {B.}~\bibnamefont {Keimer}}, \bibinfo {author}
  {\bibfnamefont {N.}~\bibnamefont {Nagaosa}},\ and\ \bibinfo {author}
  {\bibfnamefont {Y.}~\bibnamefont {Tokura}},\ }\href
  {https://doi.org/10.1038/nmat3223} {\bibfield  {journal} {\bibinfo  {journal}
  {Nat. Mater.}\ }\textbf {\bibinfo {volume} {11}},\ \bibinfo {pages} {103}
  (\bibinfo {year} {2012})}\BibitemShut {NoStop}%
\bibitem [{\citenamefont {Pines}(2016)}]{Pines_2016}%
  \BibitemOpen
  \bibfield  {author} {\bibinfo {author} {\bibfnamefont {D.}~\bibnamefont
  {Pines}},\ }\href {https://doi.org/10.1088/0034-4885/79/9/092501} {\bibfield
  {journal} {\bibinfo  {journal} {Rep. Prog. Phys.}\ }\textbf {\bibinfo
  {volume} {79}},\ \bibinfo {pages} {092501} (\bibinfo {year}
  {2016})}\BibitemShut {NoStop}%
\bibitem [{\citenamefont {Bednorz}\ and\ \citenamefont
  {M\"uller}(1986)}]{RN110}%
  \BibitemOpen
  \bibfield  {author} {\bibinfo {author} {\bibfnamefont {J.~G.}\ \bibnamefont
  {Bednorz}}\ and\ \bibinfo {author} {\bibfnamefont {K.~A.}\ \bibnamefont
  {M\"uller}},\ }\href {https://doi.org/10.1007/BF01303701} {\bibfield
  {journal} {\bibinfo  {journal} {Z. Phys. B: Condens. Matter}\ }\textbf
  {\bibinfo {volume} {64}},\ \bibinfo {pages} {189} (\bibinfo {year}
  {1986})}\BibitemShut {NoStop}%
\bibitem [{\citenamefont {Kamihara}\ \emph
  {et~al.}(2008{\natexlab{a}})\citenamefont {Kamihara}, \citenamefont
  {Watanabe}, \citenamefont {Hirano},\ and\ \citenamefont
  {Hosono}}]{doi:10.1021/ja800073m}%
  \BibitemOpen
  \bibfield  {author} {\bibinfo {author} {\bibfnamefont {Y.}~\bibnamefont
  {Kamihara}}, \bibinfo {author} {\bibfnamefont {T.}~\bibnamefont {Watanabe}},
  \bibinfo {author} {\bibfnamefont {M.}~\bibnamefont {Hirano}},\ and\ \bibinfo
  {author} {\bibfnamefont {H.}~\bibnamefont {Hosono}},\ }\href
  {https://doi.org/10.1021/ja800073m} {\bibfield  {journal} {\bibinfo
  {journal} {J. Am. Chem. Soc.}\ }\textbf {\bibinfo {volume} {130}},\ \bibinfo
  {pages} {3296} (\bibinfo {year} {2008}{\natexlab{a}})}\BibitemShut {NoStop}%
\bibitem [{\citenamefont {Keimer}\ \emph {et~al.}(2015)\citenamefont {Keimer},
  \citenamefont {Kivelson}, \citenamefont {Norman}, \citenamefont {Uchida},\
  and\ \citenamefont {Zaanen}}]{RN116}%
  \BibitemOpen
  \bibfield  {author} {\bibinfo {author} {\bibfnamefont {B.}~\bibnamefont
  {Keimer}}, \bibinfo {author} {\bibfnamefont {S.~A.}\ \bibnamefont
  {Kivelson}}, \bibinfo {author} {\bibfnamefont {M.~R.}\ \bibnamefont
  {Norman}}, \bibinfo {author} {\bibfnamefont {S.}~\bibnamefont {Uchida}},\
  and\ \bibinfo {author} {\bibfnamefont {J.}~\bibnamefont {Zaanen}},\ }\href
  {https://doi.org/10.1038/nature14165} {\bibfield  {journal} {\bibinfo
  {journal} {Nature (London)}\ }\textbf {\bibinfo {volume} {518}},\ \bibinfo
  {pages} {179} (\bibinfo {year} {2015})}\BibitemShut {NoStop}%
\bibitem [{\citenamefont {Kamihara}\ \emph
  {et~al.}(2008{\natexlab{b}})\citenamefont {Kamihara}, \citenamefont
  {Hiramatsu}, \citenamefont {Hirano}, \citenamefont {Kobayashi}, \citenamefont
  {Kitao}, \citenamefont {Higashitaniguchi}, \citenamefont {Yoda},
  \citenamefont {Seto},\ and\ \citenamefont {Hosono}}]{PhysRevB.78.184512}%
  \BibitemOpen
  \bibfield  {author} {\bibinfo {author} {\bibfnamefont {Y.}~\bibnamefont
  {Kamihara}}, \bibinfo {author} {\bibfnamefont {H.}~\bibnamefont {Hiramatsu}},
  \bibinfo {author} {\bibfnamefont {M.}~\bibnamefont {Hirano}}, \bibinfo
  {author} {\bibfnamefont {Y.}~\bibnamefont {Kobayashi}}, \bibinfo {author}
  {\bibfnamefont {S.}~\bibnamefont {Kitao}}, \bibinfo {author} {\bibfnamefont
  {S.}~\bibnamefont {Higashitaniguchi}}, \bibinfo {author} {\bibfnamefont
  {Y.}~\bibnamefont {Yoda}}, \bibinfo {author} {\bibfnamefont {M.}~\bibnamefont
  {Seto}},\ and\ \bibinfo {author} {\bibfnamefont {H.}~\bibnamefont {Hosono}},\
  }\href {https://doi.org/10.1103/PhysRevB.78.184512} {\bibfield  {journal}
  {\bibinfo  {journal} {Phys. Rev. B}\ }\textbf {\bibinfo {volume} {78}},\
  \bibinfo {pages} {184512} (\bibinfo {year} {2008}{\natexlab{b}})}\BibitemShut
  {NoStop}%
\bibitem [{\citenamefont {Lu}\ \emph {et~al.}(2015)\citenamefont {Lu},
  \citenamefont {Wang}, \citenamefont {Wu}, \citenamefont {Wu}, \citenamefont
  {Zhao}, \citenamefont {Zeng}, \citenamefont {Luo}, \citenamefont {Wu},
  \citenamefont {Bao}, \citenamefont {Zhang}, \citenamefont {Huang},
  \citenamefont {Huang},\ and\ \citenamefont {Chen}}]{RN125}%
  \BibitemOpen
  \bibfield  {author} {\bibinfo {author} {\bibfnamefont {X.~F.}\ \bibnamefont
  {Lu}}, \bibinfo {author} {\bibfnamefont {N.~Z.}\ \bibnamefont {Wang}},
  \bibinfo {author} {\bibfnamefont {H.}~\bibnamefont {Wu}}, \bibinfo {author}
  {\bibfnamefont {Y.~P.}\ \bibnamefont {Wu}}, \bibinfo {author} {\bibfnamefont
  {D.}~\bibnamefont {Zhao}}, \bibinfo {author} {\bibfnamefont {X.~Z.}\
  \bibnamefont {Zeng}}, \bibinfo {author} {\bibfnamefont {X.~G.}\ \bibnamefont
  {Luo}}, \bibinfo {author} {\bibfnamefont {T.}~\bibnamefont {Wu}}, \bibinfo
  {author} {\bibfnamefont {W.}~\bibnamefont {Bao}}, \bibinfo {author}
  {\bibfnamefont {G.~H.}\ \bibnamefont {Zhang}}, \bibinfo {author}
  {\bibfnamefont {F.~Q.}\ \bibnamefont {Huang}}, \bibinfo {author}
  {\bibfnamefont {Q.~Z.}\ \bibnamefont {Huang}},\ and\ \bibinfo {author}
  {\bibfnamefont {X.~H.}\ \bibnamefont {Chen}},\ }\href
  {https://doi.org/10.1038/nmat4155} {\bibfield  {journal} {\bibinfo  {journal}
  {Nat. Mater.}\ }\textbf {\bibinfo {volume} {14}},\ \bibinfo {pages} {325}
  (\bibinfo {year} {2015})}\BibitemShut {NoStop}%
\bibitem [{\citenamefont {Anderson}\ \emph {et~al.}(2004)\citenamefont
  {Anderson}, \citenamefont {Lee}, \citenamefont {Randeria}, \citenamefont
  {Rice}, \citenamefont {Trivedi},\ and\ \citenamefont
  {Zhang}}]{Anderson_2004}%
  \BibitemOpen
  \bibfield  {author} {\bibinfo {author} {\bibfnamefont {P.~W.}\ \bibnamefont
  {Anderson}}, \bibinfo {author} {\bibfnamefont {P.~A.}\ \bibnamefont {Lee}},
  \bibinfo {author} {\bibfnamefont {M.}~\bibnamefont {Randeria}}, \bibinfo
  {author} {\bibfnamefont {T.~M.}\ \bibnamefont {Rice}}, \bibinfo {author}
  {\bibfnamefont {N.}~\bibnamefont {Trivedi}},\ and\ \bibinfo {author}
  {\bibfnamefont {F.~C.}\ \bibnamefont {Zhang}},\ }\href
  {https://doi.org/10.1088/0953-8984/16/24/r02} {\bibfield  {journal} {\bibinfo
   {journal} {J. Phys.: Condens. Matter}\ }\textbf {\bibinfo {volume} {16}},\
  \bibinfo {pages} {R755} (\bibinfo {year} {2004})}\BibitemShut {NoStop}%
\bibitem [{\citenamefont {Seo}\ \emph {et~al.}(2008)\citenamefont {Seo},
  \citenamefont {Bernevig},\ and\ \citenamefont {Hu}}]{PhysRevLett.101.206404}%
  \BibitemOpen
  \bibfield  {author} {\bibinfo {author} {\bibfnamefont {K.}~\bibnamefont
  {Seo}}, \bibinfo {author} {\bibfnamefont {B.~A.}\ \bibnamefont {Bernevig}},\
  and\ \bibinfo {author} {\bibfnamefont {J.}~\bibnamefont {Hu}},\ }\href
  {https://doi.org/10.1103/PhysRevLett.101.206404} {\bibfield  {journal}
  {\bibinfo  {journal} {Phys. Rev. Lett.}\ }\textbf {\bibinfo {volume} {101}},\
  \bibinfo {pages} {206404} (\bibinfo {year} {2008})}\BibitemShut {NoStop}%
\bibitem [{\citenamefont {Scalapino}(2012)}]{RevModPhys.84.1383}%
  \BibitemOpen
  \bibfield  {author} {\bibinfo {author} {\bibfnamefont {D.~J.}\ \bibnamefont
  {Scalapino}},\ }\href {https://doi.org/10.1103/RevModPhys.84.1383} {\bibfield
   {journal} {\bibinfo  {journal} {Rev. Mod. Phys.}\ }\textbf {\bibinfo
  {volume} {84}},\ \bibinfo {pages} {1383} (\bibinfo {year}
  {2012})}\BibitemShut {NoStop}%
\bibitem [{\citenamefont {Dai}\ \emph {et~al.}(2012)\citenamefont {Dai},
  \citenamefont {Hu},\ and\ \citenamefont {Dagotto}}]{RN111}%
  \BibitemOpen
  \bibfield  {author} {\bibinfo {author} {\bibfnamefont {P.}~\bibnamefont
  {Dai}}, \bibinfo {author} {\bibfnamefont {J.}~\bibnamefont {Hu}},\ and\
  \bibinfo {author} {\bibfnamefont {E.}~\bibnamefont {Dagotto}},\ }\href
  {https://doi.org/10.1038/nphys2438} {\bibfield  {journal} {\bibinfo
  {journal} {Nat. Phys.}\ }\textbf {\bibinfo {volume} {8}},\ \bibinfo {pages}
  {709} (\bibinfo {year} {2012})}\BibitemShut {NoStop}%
\bibitem [{\citenamefont {Dagotto}(2013)}]{RevModPhys.85.849}%
  \BibitemOpen
  \bibfield  {author} {\bibinfo {author} {\bibfnamefont {E.}~\bibnamefont
  {Dagotto}},\ }\href {https://doi.org/10.1103/RevModPhys.85.849} {\bibfield
  {journal} {\bibinfo  {journal} {Rev. Mod. Phys.}\ }\textbf {\bibinfo {volume}
  {85}},\ \bibinfo {pages} {849} (\bibinfo {year} {2013})}\BibitemShut
  {NoStop}%
\bibitem [{\citenamefont {Okada}\ \emph {et~al.}(2017)\citenamefont {Okada},
  \citenamefont {Ando}, \citenamefont {Shimizu}, \citenamefont {Minamitani},
  \citenamefont {Shiraki}, \citenamefont {Watanabe},\ and\ \citenamefont
  {Hitosugi}}]{RN28}%
  \BibitemOpen
  \bibfield  {author} {\bibinfo {author} {\bibfnamefont {Y.}~\bibnamefont
  {Okada}}, \bibinfo {author} {\bibfnamefont {Y.}~\bibnamefont {Ando}},
  \bibinfo {author} {\bibfnamefont {R.}~\bibnamefont {Shimizu}}, \bibinfo
  {author} {\bibfnamefont {E.}~\bibnamefont {Minamitani}}, \bibinfo {author}
  {\bibfnamefont {S.}~\bibnamefont {Shiraki}}, \bibinfo {author} {\bibfnamefont
  {S.}~\bibnamefont {Watanabe}},\ and\ \bibinfo {author} {\bibfnamefont
  {T.}~\bibnamefont {Hitosugi}},\ }\href {https://doi.org/10.1038/ncomms15975}
  {\bibfield  {journal} {\bibinfo  {journal} {Nat. Commun.}\ }\textbf {\bibinfo
  {volume} {8}},\ \bibinfo {pages} {15975} (\bibinfo {year}
  {2017})}\BibitemShut {NoStop}%
\bibitem [{\citenamefont {Zhang}\ \emph {et~al.}(2019)\citenamefont {Zhang},
  \citenamefont {Fan}, \citenamefont {Chen}, \citenamefont {Wang},
  \citenamefont {Liu}, \citenamefont {Li}, \citenamefont {Yin},\ and\
  \citenamefont {Li}}]{RN114}%
  \BibitemOpen
  \bibfield  {author} {\bibinfo {author} {\bibfnamefont {C.}~\bibnamefont
  {Zhang}}, \bibinfo {author} {\bibfnamefont {Y.}~\bibnamefont {Fan}}, \bibinfo
  {author} {\bibfnamefont {Q.}~\bibnamefont {Chen}}, \bibinfo {author}
  {\bibfnamefont {T.}~\bibnamefont {Wang}}, \bibinfo {author} {\bibfnamefont
  {X.}~\bibnamefont {Liu}}, \bibinfo {author} {\bibfnamefont {Q.}~\bibnamefont
  {Li}}, \bibinfo {author} {\bibfnamefont {Y.}~\bibnamefont {Yin}},\ and\
  \bibinfo {author} {\bibfnamefont {X.}~\bibnamefont {Li}},\ }\href
  {https://doi.org/10.1038/s41427-019-0181-3} {\bibfield  {journal} {\bibinfo
  {journal} {NPG Asia Mater.}\ }\textbf {\bibinfo {volume} {11}},\ \bibinfo
  {pages} {76} (\bibinfo {year} {2019})}\BibitemShut {NoStop}%
\bibitem [{\citenamefont {Fan}\ \emph {et~al.}(2018)\citenamefont {Fan},
  \citenamefont {Ma}, \citenamefont {Wang}, \citenamefont {Zhang},
  \citenamefont {Chen}, \citenamefont {Liu}, \citenamefont {Wang},
  \citenamefont {Li}, \citenamefont {Yin},\ and\ \citenamefont
  {Li}}]{PhysRevB.98.064501}%
  \BibitemOpen
  \bibfield  {author} {\bibinfo {author} {\bibfnamefont {Y.~J.}\ \bibnamefont
  {Fan}}, \bibinfo {author} {\bibfnamefont {C.}~\bibnamefont {Ma}}, \bibinfo
  {author} {\bibfnamefont {T.~Y.}\ \bibnamefont {Wang}}, \bibinfo {author}
  {\bibfnamefont {C.}~\bibnamefont {Zhang}}, \bibinfo {author} {\bibfnamefont
  {Q.~L.}\ \bibnamefont {Chen}}, \bibinfo {author} {\bibfnamefont
  {X.}~\bibnamefont {Liu}}, \bibinfo {author} {\bibfnamefont {Z.~Q.}\
  \bibnamefont {Wang}}, \bibinfo {author} {\bibfnamefont {Q.}~\bibnamefont
  {Li}}, \bibinfo {author} {\bibfnamefont {Y.~W.}\ \bibnamefont {Yin}},\ and\
  \bibinfo {author} {\bibfnamefont {X.~G.}\ \bibnamefont {Li}},\ }\href
  {https://doi.org/10.1103/PhysRevB.98.064501} {\bibfield  {journal} {\bibinfo
  {journal} {Phys. Rev. B}\ }\textbf {\bibinfo {volume} {98}},\ \bibinfo
  {pages} {064501} (\bibinfo {year} {2018})}\BibitemShut {NoStop}%
\bibitem [{\citenamefont {Jia}\ \emph {et~al.}(2018)\citenamefont {Jia},
  \citenamefont {He}, \citenamefont {Hu}, \citenamefont {Yang}, \citenamefont
  {Yang}, \citenamefont {Yu}, \citenamefont {Zhang}, \citenamefont {Shi},
  \citenamefont {Lin}, \citenamefont {Yuan}, \citenamefont {Zhu}, \citenamefont
  {Gu}, \citenamefont {Li},\ and\ \citenamefont {Jin}}]{RN34}%
  \BibitemOpen
  \bibfield  {author} {\bibinfo {author} {\bibfnamefont {Y.}~\bibnamefont
  {Jia}}, \bibinfo {author} {\bibfnamefont {G.}~\bibnamefont {He}}, \bibinfo
  {author} {\bibfnamefont {W.}~\bibnamefont {Hu}}, \bibinfo {author}
  {\bibfnamefont {H.}~\bibnamefont {Yang}}, \bibinfo {author} {\bibfnamefont
  {Z.}~\bibnamefont {Yang}}, \bibinfo {author} {\bibfnamefont {H.}~\bibnamefont
  {Yu}}, \bibinfo {author} {\bibfnamefont {Q.}~\bibnamefont {Zhang}}, \bibinfo
  {author} {\bibfnamefont {J.}~\bibnamefont {Shi}}, \bibinfo {author}
  {\bibfnamefont {Z.}~\bibnamefont {Lin}}, \bibinfo {author} {\bibfnamefont
  {J.}~\bibnamefont {Yuan}}, \bibinfo {author} {\bibfnamefont {B.}~\bibnamefont
  {Zhu}}, \bibinfo {author} {\bibfnamefont {L.}~\bibnamefont {Gu}}, \bibinfo
  {author} {\bibfnamefont {H.}~\bibnamefont {Li}},\ and\ \bibinfo {author}
  {\bibfnamefont {K.}~\bibnamefont {Jin}},\ }\href
  {https://doi.org/10.1038/s41598-018-22393-8} {\bibfield  {journal} {\bibinfo
  {journal} {Sci. Rep.}\ }\textbf {\bibinfo {volume} {8}},\ \bibinfo {pages}
  {3995} (\bibinfo {year} {2018})}\BibitemShut {NoStop}%
\bibitem [{\citenamefont {Wei}\ \emph {et~al.}(2021)\citenamefont {Wei},
  \citenamefont {Li}, \citenamefont {Gong}, \citenamefont {Wei}, \citenamefont
  {Hu}, \citenamefont {Ni}, \citenamefont {He}, \citenamefont {Qin},
  \citenamefont {Kusmartseva}, \citenamefont {Kusmartsev}, \citenamefont
  {Yuan}, \citenamefont {Zhu}, \citenamefont {Chen}, \citenamefont {Chen},
  \citenamefont {Liu},\ and\ \citenamefont {Jin}}]{PhysRevB.103.L140501}%
  \BibitemOpen
  \bibfield  {author} {\bibinfo {author} {\bibfnamefont {Z.}~\bibnamefont
  {Wei}}, \bibinfo {author} {\bibfnamefont {Q.}~\bibnamefont {Li}}, \bibinfo
  {author} {\bibfnamefont {B.-C.}\ \bibnamefont {Gong}}, \bibinfo {author}
  {\bibfnamefont {X.}~\bibnamefont {Wei}}, \bibinfo {author} {\bibfnamefont
  {W.}~\bibnamefont {Hu}}, \bibinfo {author} {\bibfnamefont {Z.}~\bibnamefont
  {Ni}}, \bibinfo {author} {\bibfnamefont {G.}~\bibnamefont {He}}, \bibinfo
  {author} {\bibfnamefont {M.}~\bibnamefont {Qin}}, \bibinfo {author}
  {\bibfnamefont {A.}~\bibnamefont {Kusmartseva}}, \bibinfo {author}
  {\bibfnamefont {F.~V.}\ \bibnamefont {Kusmartsev}}, \bibinfo {author}
  {\bibfnamefont {J.}~\bibnamefont {Yuan}}, \bibinfo {author} {\bibfnamefont
  {B.}~\bibnamefont {Zhu}}, \bibinfo {author} {\bibfnamefont {Q.}~\bibnamefont
  {Chen}}, \bibinfo {author} {\bibfnamefont {J.-H.}\ \bibnamefont {Chen}},
  \bibinfo {author} {\bibfnamefont {K.}~\bibnamefont {Liu}},\ and\ \bibinfo
  {author} {\bibfnamefont {K.}~\bibnamefont {Jin}},\ }\href
  {https://doi.org/10.1103/PhysRevB.103.L140501} {\bibfield  {journal}
  {\bibinfo  {journal} {Phys. Rev. B}\ }\textbf {\bibinfo {volume} {103}},\
  \bibinfo {pages} {L140501} (\bibinfo {year} {2021})}\BibitemShut {NoStop}%
\bibitem [{\citenamefont {Johnston}\ \emph {et~al.}(1973)\citenamefont
  {Johnston}, \citenamefont {Prakash}, \citenamefont {Zachariasen},\ and\
  \citenamefont {Viswanathan}}]{JOHNSTON1973777}%
  \BibitemOpen
  \bibfield  {author} {\bibinfo {author} {\bibfnamefont {D.~C.}\ \bibnamefont
  {Johnston}}, \bibinfo {author} {\bibfnamefont {H.}~\bibnamefont {Prakash}},
  \bibinfo {author} {\bibfnamefont {W.~H.}\ \bibnamefont {Zachariasen}},\ and\
  \bibinfo {author} {\bibfnamefont {R.}~\bibnamefont {Viswanathan}},\ }\href
  {https://doi.org/10.1016/0025-5408(73)90183-9} {\bibfield  {journal}
  {\bibinfo  {journal} {Mater. Res. Bull.}\ }\textbf {\bibinfo {volume} {8}},\
  \bibinfo {pages} {777} (\bibinfo {year} {1973})}\BibitemShut {NoStop}%
\bibitem [{\citenamefont {Wang}\ \emph {et~al.}(2017)\citenamefont {Wang},
  \citenamefont {Huang}, \citenamefont {He}, \citenamefont {Che}, \citenamefont
  {Zhang},\ and\ \citenamefont {Huang}}]{doi:10.1021/acsomega.7b00048}%
  \BibitemOpen
  \bibfield  {author} {\bibinfo {author} {\bibfnamefont {D.}~\bibnamefont
  {Wang}}, \bibinfo {author} {\bibfnamefont {C.}~\bibnamefont {Huang}},
  \bibinfo {author} {\bibfnamefont {J.}~\bibnamefont {He}}, \bibinfo {author}
  {\bibfnamefont {X.}~\bibnamefont {Che}}, \bibinfo {author} {\bibfnamefont
  {H.}~\bibnamefont {Zhang}},\ and\ \bibinfo {author} {\bibfnamefont
  {F.}~\bibnamefont {Huang}},\ }\href
  {https://doi.org/10.1021/acsomega.7b00048} {\bibfield  {journal} {\bibinfo
  {journal} {ACS Omega}\ }\textbf {\bibinfo {volume} {2}},\ \bibinfo {pages}
  {1036} (\bibinfo {year} {2017})}\BibitemShut {NoStop}%
\bibitem [{\citenamefont {Yoshimatsu}\ \emph {et~al.}(2017)\citenamefont
  {Yoshimatsu}, \citenamefont {Sakata},\ and\ \citenamefont {Ohtomo}}]{RN112}%
  \BibitemOpen
  \bibfield  {author} {\bibinfo {author} {\bibfnamefont {K.}~\bibnamefont
  {Yoshimatsu}}, \bibinfo {author} {\bibfnamefont {O.}~\bibnamefont {Sakata}},\
  and\ \bibinfo {author} {\bibfnamefont {A.}~\bibnamefont {Ohtomo}},\ }\href
  {https://doi.org/10.1038/s41598-017-12815-4} {\bibfield  {journal} {\bibinfo
  {journal} {Sci. Rep.}\ }\textbf {\bibinfo {volume} {7}},\ \bibinfo {pages}
  {12544} (\bibinfo {year} {2017})}\BibitemShut {NoStop}%
\bibitem [{\citenamefont {Massidda}\ \emph {et~al.}(1988)\citenamefont
  {Massidda}, \citenamefont {Yu},\ and\ \citenamefont
  {Freeman}}]{PhysRevB.38.11352}%
  \BibitemOpen
  \bibfield  {author} {\bibinfo {author} {\bibfnamefont {S.}~\bibnamefont
  {Massidda}}, \bibinfo {author} {\bibfnamefont {J.}~\bibnamefont {Yu}},\ and\
  \bibinfo {author} {\bibfnamefont {A.~J.}\ \bibnamefont {Freeman}},\ }\href
  {https://doi.org/10.1103/PhysRevB.38.11352} {\bibfield  {journal} {\bibinfo
  {journal} {Phys. Rev. B}\ }\textbf {\bibinfo {volume} {38}},\ \bibinfo
  {pages} {11352} (\bibinfo {year} {1988})}\BibitemShut {NoStop}%
\bibitem [{\citenamefont {Chen}\ \emph {et~al.}(2011)\citenamefont {Chen},
  \citenamefont {Dong}, \citenamefont {Asokan}, \citenamefont {Chen},
  \citenamefont {Liu}, \citenamefont {Guo}, \citenamefont {Yang}, \citenamefont
  {Chen}, \citenamefont {Hsu}, \citenamefont {Chang},\ and\ \citenamefont
  {Wu}}]{Chen_2011}%
  \BibitemOpen
  \bibfield  {author} {\bibinfo {author} {\bibfnamefont {C.~L.}\ \bibnamefont
  {Chen}}, \bibinfo {author} {\bibfnamefont {C.~L.}\ \bibnamefont {Dong}},
  \bibinfo {author} {\bibfnamefont {K.}~\bibnamefont {Asokan}}, \bibinfo
  {author} {\bibfnamefont {J.~L.}\ \bibnamefont {Chen}}, \bibinfo {author}
  {\bibfnamefont {Y.~S.}\ \bibnamefont {Liu}}, \bibinfo {author} {\bibfnamefont
  {J.-H.}\ \bibnamefont {Guo}}, \bibinfo {author} {\bibfnamefont {W.~L.}\
  \bibnamefont {Yang}}, \bibinfo {author} {\bibfnamefont {Y.~Y.}\ \bibnamefont
  {Chen}}, \bibinfo {author} {\bibfnamefont {F.~C.}\ \bibnamefont {Hsu}},
  \bibinfo {author} {\bibfnamefont {C.~L.}\ \bibnamefont {Chang}},\ and\
  \bibinfo {author} {\bibfnamefont {M.~K.}\ \bibnamefont {Wu}},\ }\href
  {https://doi.org/10.1088/0953-2048/24/11/115007} {\bibfield  {journal}
  {\bibinfo  {journal} {Supercond. Sci. Technol.}\ }\textbf {\bibinfo {volume}
  {24}},\ \bibinfo {pages} {115007} (\bibinfo {year} {2011})}\BibitemShut
  {NoStop}%
\bibitem [{\citenamefont {Zhang}\ \emph {et~al.}(2017)\citenamefont {Zhang},
  \citenamefont {Hao}, \citenamefont {Gao}, \citenamefont {Liu}, \citenamefont
  {Ma}, \citenamefont {Lin}, \citenamefont {Yin},\ and\ \citenamefont
  {Li}}]{zhang2017enhanced}%
  \BibitemOpen
  \bibfield  {author} {\bibinfo {author} {\bibfnamefont {C.}~\bibnamefont
  {Zhang}}, \bibinfo {author} {\bibfnamefont {F.}~\bibnamefont {Hao}}, \bibinfo
  {author} {\bibfnamefont {G.}~\bibnamefont {Gao}}, \bibinfo {author}
  {\bibfnamefont {X.}~\bibnamefont {Liu}}, \bibinfo {author} {\bibfnamefont
  {C.}~\bibnamefont {Ma}}, \bibinfo {author} {\bibfnamefont {Y.}~\bibnamefont
  {Lin}}, \bibinfo {author} {\bibfnamefont {Y.}~\bibnamefont {Yin}},\ and\
  \bibinfo {author} {\bibfnamefont {X.}~\bibnamefont {Li}},\ }\href
  {https://www.nature.com/articles/s41535-016-0006-3} {\bibfield  {journal}
  {\bibinfo  {journal} {npj Quantum Mater.}\ }\textbf {\bibinfo {volume} {2}},\
  \bibinfo {pages} {2} (\bibinfo {year} {2017})}\BibitemShut {NoStop}%
\bibitem [{\citenamefont {Fan}\ \emph {et~al.}(2019)\citenamefont {Fan},
  \citenamefont {Zhang}, \citenamefont {Liu}, \citenamefont {Lin},
  \citenamefont {Gao}, \citenamefont {Ma}, \citenamefont {Yin},\ and\
  \citenamefont {Li}}]{FAN2019607}%
  \BibitemOpen
  \bibfield  {author} {\bibinfo {author} {\bibfnamefont {Y.}~\bibnamefont
  {Fan}}, \bibinfo {author} {\bibfnamefont {C.}~\bibnamefont {Zhang}}, \bibinfo
  {author} {\bibfnamefont {X.}~\bibnamefont {Liu}}, \bibinfo {author}
  {\bibfnamefont {Y.}~\bibnamefont {Lin}}, \bibinfo {author} {\bibfnamefont
  {G.}~\bibnamefont {Gao}}, \bibinfo {author} {\bibfnamefont {C.}~\bibnamefont
  {Ma}}, \bibinfo {author} {\bibfnamefont {Y.}~\bibnamefont {Yin}},\ and\
  \bibinfo {author} {\bibfnamefont {X.}~\bibnamefont {Li}},\ }\href
  {https://doi.org/10.1016/j.jallcom.2019.01.381} {\bibfield  {journal}
  {\bibinfo  {journal} {J. Alloys Compd.}\ }\textbf {\bibinfo {volume} {786}},\
  \bibinfo {pages} {607} (\bibinfo {year} {2019})}\BibitemShut {NoStop}%
\bibitem [{\citenamefont {Li}\ \emph {et~al.}(2021)\citenamefont {Li},
  \citenamefont {Zou}, \citenamefont {Han}, \citenamefont {Foyevtsova},
  \citenamefont {Shin}, \citenamefont {Lee}, \citenamefont {Liu}, \citenamefont
  {Shin}, \citenamefont {Albright}, \citenamefont {Sutarto}, \citenamefont
  {He}, \citenamefont {Davidson}, \citenamefont {Walker}, \citenamefont {Ahn},
  \citenamefont {Zhu}, \citenamefont {Cheng}, \citenamefont {Elfimov},
  \citenamefont {Sawatzky},\ and\ \citenamefont
  {Zou}}]{doi:10.1126/sciadv.abd4248}%
  \BibitemOpen
  \bibfield  {author} {\bibinfo {author} {\bibfnamefont {F.}~\bibnamefont
  {Li}}, \bibinfo {author} {\bibfnamefont {Y.}~\bibnamefont {Zou}}, \bibinfo
  {author} {\bibfnamefont {M.-G.}\ \bibnamefont {Han}}, \bibinfo {author}
  {\bibfnamefont {K.}~\bibnamefont {Foyevtsova}}, \bibinfo {author}
  {\bibfnamefont {H.}~\bibnamefont {Shin}}, \bibinfo {author} {\bibfnamefont
  {S.}~\bibnamefont {Lee}}, \bibinfo {author} {\bibfnamefont {C.}~\bibnamefont
  {Liu}}, \bibinfo {author} {\bibfnamefont {K.}~\bibnamefont {Shin}}, \bibinfo
  {author} {\bibfnamefont {S.~D.}\ \bibnamefont {Albright}}, \bibinfo {author}
  {\bibfnamefont {R.}~\bibnamefont {Sutarto}}, \bibinfo {author} {\bibfnamefont
  {F.}~\bibnamefont {He}}, \bibinfo {author} {\bibfnamefont {B.~A.}\
  \bibnamefont {Davidson}}, \bibinfo {author} {\bibfnamefont {F.~J.}\
  \bibnamefont {Walker}}, \bibinfo {author} {\bibfnamefont {C.~H.}\
  \bibnamefont {Ahn}}, \bibinfo {author} {\bibfnamefont {Y.}~\bibnamefont
  {Zhu}}, \bibinfo {author} {\bibfnamefont {Z.~G.}\ \bibnamefont {Cheng}},
  \bibinfo {author} {\bibfnamefont {I.}~\bibnamefont {Elfimov}}, \bibinfo
  {author} {\bibfnamefont {G.~A.}\ \bibnamefont {Sawatzky}},\ and\ \bibinfo
  {author} {\bibfnamefont {K.}~\bibnamefont {Zou}},\ }\href
  {https://doi.org/10.1126/sciadv.abd4248} {\bibfield  {journal} {\bibinfo
  {journal} {Sci. Adv.}\ }\textbf {\bibinfo {volume} {7}},\ \bibinfo {pages}
  {eabd4248} (\bibinfo {year} {2021})}\BibitemShut {NoStop}%
\bibitem [{\citenamefont {Jin}\ \emph {et~al.}(2015)\citenamefont {Jin},
  \citenamefont {He}, \citenamefont {Zhang}, \citenamefont {Maruyama},
  \citenamefont {Yasui}, \citenamefont {Suchoski}, \citenamefont {Shin},
  \citenamefont {Jiang}, \citenamefont {Yu}, \citenamefont {Yuan},
  \citenamefont {Shan}, \citenamefont {Kusmartsev}, \citenamefont {Greene},\
  and\ \citenamefont {Takeuchi}}]{RN3}%
  \BibitemOpen
  \bibfield  {author} {\bibinfo {author} {\bibfnamefont {K.}~\bibnamefont
  {Jin}}, \bibinfo {author} {\bibfnamefont {G.}~\bibnamefont {He}}, \bibinfo
  {author} {\bibfnamefont {X.}~\bibnamefont {Zhang}}, \bibinfo {author}
  {\bibfnamefont {S.}~\bibnamefont {Maruyama}}, \bibinfo {author}
  {\bibfnamefont {S.}~\bibnamefont {Yasui}}, \bibinfo {author} {\bibfnamefont
  {R.}~\bibnamefont {Suchoski}}, \bibinfo {author} {\bibfnamefont
  {J.}~\bibnamefont {Shin}}, \bibinfo {author} {\bibfnamefont {Y.}~\bibnamefont
  {Jiang}}, \bibinfo {author} {\bibfnamefont {H.~S.}\ \bibnamefont {Yu}},
  \bibinfo {author} {\bibfnamefont {J.}~\bibnamefont {Yuan}}, \bibinfo {author}
  {\bibfnamefont {L.}~\bibnamefont {Shan}}, \bibinfo {author} {\bibfnamefont
  {F.~V.}\ \bibnamefont {Kusmartsev}}, \bibinfo {author} {\bibfnamefont
  {R.~L.}\ \bibnamefont {Greene}},\ and\ \bibinfo {author} {\bibfnamefont
  {I.}~\bibnamefont {Takeuchi}},\ }\href {https://doi.org/10.1038/ncomms8183}
  {\bibfield  {journal} {\bibinfo  {journal} {Nat. Commun.}\ }\textbf {\bibinfo
  {volume} {6}},\ \bibinfo {pages} {7183} (\bibinfo {year} {2015})}\BibitemShut
  {NoStop}%
\bibitem [{\citenamefont {Fan}\ \emph {et~al.}(2020)\citenamefont {Fan},
  \citenamefont {Gan}, \citenamefont {Wang}, \citenamefont {Sun}, \citenamefont
  {Ma}, \citenamefont {Huang}, \citenamefont {Zhou}, \citenamefont {Yin},\ and\
  \citenamefont {Li}}]{FAN202066}%
  \BibitemOpen
  \bibfield  {author} {\bibinfo {author} {\bibfnamefont {Y.~J.}\ \bibnamefont
  {Fan}}, \bibinfo {author} {\bibfnamefont {H.}~\bibnamefont {Gan}}, \bibinfo
  {author} {\bibfnamefont {D.}~\bibnamefont {Wang}}, \bibinfo {author}
  {\bibfnamefont {H.~Y.}\ \bibnamefont {Sun}}, \bibinfo {author} {\bibfnamefont
  {C.}~\bibnamefont {Ma}}, \bibinfo {author} {\bibfnamefont {F.~Q.}\
  \bibnamefont {Huang}}, \bibinfo {author} {\bibfnamefont {J.}~\bibnamefont
  {Zhou}}, \bibinfo {author} {\bibfnamefont {Y.~W.}\ \bibnamefont {Yin}},\ and\
  \bibinfo {author} {\bibfnamefont {X.}~\bibnamefont {Li}},\ }\href
  {https://doi.org/10.1016/j.actamat.2020.09.001} {\bibfield  {journal}
  {\bibinfo  {journal} {Acta Mater.}\ }\textbf {\bibinfo {volume} {200}},\
  \bibinfo {pages} {66} (\bibinfo {year} {2020})}\BibitemShut {NoStop}%
\bibitem [{\citenamefont {Luo}\ and\ \citenamefont {Chen}(2015)}]{RN115}%
  \BibitemOpen
  \bibfield  {author} {\bibinfo {author} {\bibfnamefont {X.}~\bibnamefont
  {Luo}}\ and\ \bibinfo {author} {\bibfnamefont {X.}~\bibnamefont {Chen}},\
  }\href {https://doi.org/10.1007/s40843-015-0022-9} {\bibfield  {journal}
  {\bibinfo  {journal} {Sci. China Mater.}\ }\textbf {\bibinfo {volume} {58}},\
  \bibinfo {pages} {77} (\bibinfo {year} {2015})}\BibitemShut {NoStop}%
\bibitem [{\citenamefont {Hu}\ \emph {et~al.}(2020)\citenamefont {Hu},
  \citenamefont {Feng}, \citenamefont {Gong}, \citenamefont {He}, \citenamefont
  {Li}, \citenamefont {Qin}, \citenamefont {Shi}, \citenamefont {Li},
  \citenamefont {Zhang}, \citenamefont {Yuan}, \citenamefont {Zhu},
  \citenamefont {Liu}, \citenamefont {Xiang}, \citenamefont {Gu}, \citenamefont
  {Zhou}, \citenamefont {Dong}, \citenamefont {Zhao},\ and\ \citenamefont
  {Jin}}]{PhysRevB.101.220510}%
  \BibitemOpen
  \bibfield  {author} {\bibinfo {author} {\bibfnamefont {W.}~\bibnamefont
  {Hu}}, \bibinfo {author} {\bibfnamefont {Z.}~\bibnamefont {Feng}}, \bibinfo
  {author} {\bibfnamefont {B.-C.}\ \bibnamefont {Gong}}, \bibinfo {author}
  {\bibfnamefont {G.}~\bibnamefont {He}}, \bibinfo {author} {\bibfnamefont
  {D.}~\bibnamefont {Li}}, \bibinfo {author} {\bibfnamefont {M.}~\bibnamefont
  {Qin}}, \bibinfo {author} {\bibfnamefont {Y.}~\bibnamefont {Shi}}, \bibinfo
  {author} {\bibfnamefont {Q.}~\bibnamefont {Li}}, \bibinfo {author}
  {\bibfnamefont {Q.}~\bibnamefont {Zhang}}, \bibinfo {author} {\bibfnamefont
  {J.}~\bibnamefont {Yuan}}, \bibinfo {author} {\bibfnamefont {B.}~\bibnamefont
  {Zhu}}, \bibinfo {author} {\bibfnamefont {K.}~\bibnamefont {Liu}}, \bibinfo
  {author} {\bibfnamefont {T.}~\bibnamefont {Xiang}}, \bibinfo {author}
  {\bibfnamefont {L.}~\bibnamefont {Gu}}, \bibinfo {author} {\bibfnamefont
  {F.}~\bibnamefont {Zhou}}, \bibinfo {author} {\bibfnamefont {X.}~\bibnamefont
  {Dong}}, \bibinfo {author} {\bibfnamefont {Z.}~\bibnamefont {Zhao}},\ and\
  \bibinfo {author} {\bibfnamefont {K.}~\bibnamefont {Jin}},\ }\href
  {https://doi.org/10.1103/PhysRevB.101.220510} {\bibfield  {journal} {\bibinfo
   {journal} {Phys. Rev. B}\ }\textbf {\bibinfo {volume} {101}},\ \bibinfo
  {pages} {220510(R)} (\bibinfo {year} {2020})}\BibitemShut {NoStop}%
\bibitem [{\citenamefont {Schmidt}\ \emph {et~al.}(2004)\citenamefont
  {Schmidt}, \citenamefont {Ratcliff}, \citenamefont {Radaelli}, \citenamefont
  {Refson}, \citenamefont {Harrison},\ and\ \citenamefont
  {Cheong}}]{PhysRevLett.92.056402}%
  \BibitemOpen
  \bibfield  {author} {\bibinfo {author} {\bibfnamefont {M.}~\bibnamefont
  {Schmidt}}, \bibinfo {author} {\bibfnamefont {W.}~\bibnamefont {Ratcliff}},
  \bibinfo {author} {\bibfnamefont {P.~G.}\ \bibnamefont {Radaelli}}, \bibinfo
  {author} {\bibfnamefont {K.}~\bibnamefont {Refson}}, \bibinfo {author}
  {\bibfnamefont {N.~M.}\ \bibnamefont {Harrison}},\ and\ \bibinfo {author}
  {\bibfnamefont {S.~W.}\ \bibnamefont {Cheong}},\ }\href
  {https://doi.org/10.1103/PhysRevLett.92.056402} {\bibfield  {journal}
  {\bibinfo  {journal} {Phys. Rev. Lett.}\ }\textbf {\bibinfo {volume} {92}},\
  \bibinfo {pages} {056402} (\bibinfo {year} {2004})}\BibitemShut {NoStop}%
\bibitem [{\citenamefont {Di~Matteo}\ \emph {et~al.}(2004)\citenamefont
  {Di~Matteo}, \citenamefont {Jackeli}, \citenamefont {Lacroix},\ and\
  \citenamefont {Perkins}}]{PhysRevLett.93.077208}%
  \BibitemOpen
  \bibfield  {author} {\bibinfo {author} {\bibfnamefont {S.}~\bibnamefont
  {Di~Matteo}}, \bibinfo {author} {\bibfnamefont {G.}~\bibnamefont {Jackeli}},
  \bibinfo {author} {\bibfnamefont {C.}~\bibnamefont {Lacroix}},\ and\ \bibinfo
  {author} {\bibfnamefont {N.~B.}\ \bibnamefont {Perkins}},\ }\href
  {https://doi.org/10.1103/PhysRevLett.93.077208} {\bibfield  {journal}
  {\bibinfo  {journal} {Phys. Rev. Lett.}\ }\textbf {\bibinfo {volume} {93}},\
  \bibinfo {pages} {077208} (\bibinfo {year} {2004})}\BibitemShut {NoStop}%
\bibitem [{\citenamefont {Khomskii}\ and\ \citenamefont
  {Mizokawa}(2005)}]{PhysRevLett.94.156402}%
  \BibitemOpen
  \bibfield  {author} {\bibinfo {author} {\bibfnamefont {D.~I.}\ \bibnamefont
  {Khomskii}}\ and\ \bibinfo {author} {\bibfnamefont {T.}~\bibnamefont
  {Mizokawa}},\ }\href {https://doi.org/10.1103/PhysRevLett.94.156402}
  {\bibfield  {journal} {\bibinfo  {journal} {Phys. Rev. Lett.}\ }\textbf
  {\bibinfo {volume} {94}},\ \bibinfo {pages} {156402} (\bibinfo {year}
  {2005})}\BibitemShut {NoStop}%
\bibitem [{\citenamefont {Leoni}\ \emph {et~al.}(2008)\citenamefont {Leoni},
  \citenamefont {Yaresko}, \citenamefont {Perkins}, \citenamefont {Rosner},\
  and\ \citenamefont {Craco}}]{PhysRevB.78.125105}%
  \BibitemOpen
  \bibfield  {author} {\bibinfo {author} {\bibfnamefont {S.}~\bibnamefont
  {Leoni}}, \bibinfo {author} {\bibfnamefont {A.~N.}\ \bibnamefont {Yaresko}},
  \bibinfo {author} {\bibfnamefont {N.}~\bibnamefont {Perkins}}, \bibinfo
  {author} {\bibfnamefont {H.}~\bibnamefont {Rosner}},\ and\ \bibinfo {author}
  {\bibfnamefont {L.}~\bibnamefont {Craco}},\ }\href
  {https://doi.org/10.1103/PhysRevB.78.125105} {\bibfield  {journal} {\bibinfo
  {journal} {Phys. Rev. B}\ }\textbf {\bibinfo {volume} {78}},\ \bibinfo
  {pages} {125105} (\bibinfo {year} {2008})}\BibitemShut {NoStop}%
\bibitem [{\citenamefont
  {Moshopoulou}(1999)}]{10.1111/j.1151-2916.1999.tb02245.x}%
  \BibitemOpen
  \bibfield  {author} {\bibinfo {author} {\bibfnamefont {E.~G.}\ \bibnamefont
  {Moshopoulou}},\ }\href {https://doi.org/10.1111/j.1151-2916.1999.tb02245.x}
  {\bibfield  {journal} {\bibinfo  {journal} {J. Am. Ceram. Soc.}\ }\textbf
  {\bibinfo {volume} {82}},\ \bibinfo {pages} {3317} (\bibinfo {year}
  {1999})}\BibitemShut {NoStop}%
\bibitem [{\citenamefont {Gorbenko}\ \emph {et~al.}(2002)\citenamefont
  {Gorbenko}, \citenamefont {Samoilenkov}, \citenamefont {Graboy},\ and\
  \citenamefont {Kaul}}]{doi:10.1021/cm021111v}%
  \BibitemOpen
  \bibfield  {author} {\bibinfo {author} {\bibfnamefont {O.~Y.}\ \bibnamefont
  {Gorbenko}}, \bibinfo {author} {\bibfnamefont {S.~V.}\ \bibnamefont
  {Samoilenkov}}, \bibinfo {author} {\bibfnamefont {I.~E.}\ \bibnamefont
  {Graboy}},\ and\ \bibinfo {author} {\bibfnamefont {A.~R.}\ \bibnamefont
  {Kaul}},\ }\href {https://doi.org/10.1021/cm021111v} {\bibfield  {journal}
  {\bibinfo  {journal} {Chem. Mater.}\ }\textbf {\bibinfo {volume} {14}},\
  \bibinfo {pages} {4026} (\bibinfo {year} {2002})}\BibitemShut {NoStop}%
\bibitem [{\citenamefont {Jeen}\ \emph {et~al.}(2013)\citenamefont {Jeen},
  \citenamefont {Choi}, \citenamefont {Biegalski}, \citenamefont {Folkman},
  \citenamefont {Tung}, \citenamefont {Fong}, \citenamefont {Freeland},
  \citenamefont {Shin}, \citenamefont {Ohta}, \citenamefont {Chisholm},\ and\
  \citenamefont {Lee}}]{RN122}%
  \BibitemOpen
  \bibfield  {author} {\bibinfo {author} {\bibfnamefont {H.}~\bibnamefont
  {Jeen}}, \bibinfo {author} {\bibfnamefont {W.~S.}\ \bibnamefont {Choi}},
  \bibinfo {author} {\bibfnamefont {M.~D.}\ \bibnamefont {Biegalski}}, \bibinfo
  {author} {\bibfnamefont {C.~M.}\ \bibnamefont {Folkman}}, \bibinfo {author}
  {\bibfnamefont {I.~C.}\ \bibnamefont {Tung}}, \bibinfo {author}
  {\bibfnamefont {D.~D.}\ \bibnamefont {Fong}}, \bibinfo {author}
  {\bibfnamefont {J.~W.}\ \bibnamefont {Freeland}}, \bibinfo {author}
  {\bibfnamefont {D.}~\bibnamefont {Shin}}, \bibinfo {author} {\bibfnamefont
  {H.}~\bibnamefont {Ohta}}, \bibinfo {author} {\bibfnamefont {M.~F.}\
  \bibnamefont {Chisholm}},\ and\ \bibinfo {author} {\bibfnamefont {H.~N.}\
  \bibnamefont {Lee}},\ }\href {https://doi.org/10.1038/nmat3736} {\bibfield
  {journal} {\bibinfo  {journal} {Nat. Mater.}\ }\textbf {\bibinfo {volume}
  {12}},\ \bibinfo {pages} {1057} (\bibinfo {year} {2013})}\BibitemShut
  {NoStop}%
\bibitem [{\citenamefont {Ramesh}\ and\ \citenamefont
  {Schlom}(2019)}]{Ramesh2019NRM}%
  \BibitemOpen
  \bibfield  {author} {\bibinfo {author} {\bibfnamefont {R.}~\bibnamefont
  {Ramesh}}\ and\ \bibinfo {author} {\bibfnamefont {D.~G.}\ \bibnamefont
  {Schlom}},\ }\href {https://doi.org/10.1038/s41578-019-0095-2} {\bibfield
  {journal} {\bibinfo  {journal} {Nat. Rev. Mater.}\ }\textbf {\bibinfo
  {volume} {4}},\ \bibinfo {pages} {257} (\bibinfo {year} {2019})}\BibitemShut
  {NoStop}%
\bibitem [{\citenamefont {Paull}\ \emph {et~al.}(2022)\citenamefont {Paull},
  \citenamefont {Xu}, \citenamefont {Cheng}, \citenamefont {Zhang},
  \citenamefont {Xu}, \citenamefont {Kelley}, \citenamefont {de~Marco},
  \citenamefont {Vasudevan}, \citenamefont {Bellaiche}, \citenamefont
  {Nagarajan},\ and\ \citenamefont {Sando}}]{RN123}%
  \BibitemOpen
  \bibfield  {author} {\bibinfo {author} {\bibfnamefont {O.}~\bibnamefont
  {Paull}}, \bibinfo {author} {\bibfnamefont {C.}~\bibnamefont {Xu}}, \bibinfo
  {author} {\bibfnamefont {X.}~\bibnamefont {Cheng}}, \bibinfo {author}
  {\bibfnamefont {Y.}~\bibnamefont {Zhang}}, \bibinfo {author} {\bibfnamefont
  {B.}~\bibnamefont {Xu}}, \bibinfo {author} {\bibfnamefont {K.~P.}\
  \bibnamefont {Kelley}}, \bibinfo {author} {\bibfnamefont {A.}~\bibnamefont
  {de~Marco}}, \bibinfo {author} {\bibfnamefont {R.~K.}\ \bibnamefont
  {Vasudevan}}, \bibinfo {author} {\bibfnamefont {L.}~\bibnamefont
  {Bellaiche}}, \bibinfo {author} {\bibfnamefont {V.}~\bibnamefont
  {Nagarajan}},\ and\ \bibinfo {author} {\bibfnamefont {D.}~\bibnamefont
  {Sando}},\ }\href {https://doi.org/10.1038/s41563-021-01098-w} {\bibfield
  {journal} {\bibinfo  {journal} {Nat. Mater.}\ }\textbf {\bibinfo {volume}
  {21}},\ \bibinfo {pages} {74} (\bibinfo {year} {2022})}\BibitemShut {NoStop}%
\bibitem [{\citenamefont {Ruf}\ \emph {et~al.}(2021)\citenamefont {Ruf},
  \citenamefont {Paik}, \citenamefont {Schreiber}, \citenamefont {Nair},
  \citenamefont {Miao}, \citenamefont {Kawasaki}, \citenamefont {Nelson},
  \citenamefont {Faeth}, \citenamefont {Lee}, \citenamefont {Goodge},
  \citenamefont {Pamuk}, \citenamefont {Fennie}, \citenamefont {Kourkoutis},
  \citenamefont {Schlom},\ and\ \citenamefont {Shen}}]{RN124}%
  \BibitemOpen
  \bibfield  {author} {\bibinfo {author} {\bibfnamefont {J.~P.}\ \bibnamefont
  {Ruf}}, \bibinfo {author} {\bibfnamefont {H.}~\bibnamefont {Paik}}, \bibinfo
  {author} {\bibfnamefont {N.~J.}\ \bibnamefont {Schreiber}}, \bibinfo {author}
  {\bibfnamefont {H.~P.}\ \bibnamefont {Nair}}, \bibinfo {author}
  {\bibfnamefont {L.}~\bibnamefont {Miao}}, \bibinfo {author} {\bibfnamefont
  {J.~K.}\ \bibnamefont {Kawasaki}}, \bibinfo {author} {\bibfnamefont {J.~N.}\
  \bibnamefont {Nelson}}, \bibinfo {author} {\bibfnamefont {B.~D.}\
  \bibnamefont {Faeth}}, \bibinfo {author} {\bibfnamefont {Y.}~\bibnamefont
  {Lee}}, \bibinfo {author} {\bibfnamefont {B.~H.}\ \bibnamefont {Goodge}},
  \bibinfo {author} {\bibfnamefont {B.}~\bibnamefont {Pamuk}}, \bibinfo
  {author} {\bibfnamefont {C.~J.}\ \bibnamefont {Fennie}}, \bibinfo {author}
  {\bibfnamefont {L.~F.}\ \bibnamefont {Kourkoutis}}, \bibinfo {author}
  {\bibfnamefont {D.~G.}\ \bibnamefont {Schlom}},\ and\ \bibinfo {author}
  {\bibfnamefont {K.~M.}\ \bibnamefont {Shen}},\ }\href
  {https://doi.org/10.1038/s41467-020-20252-7} {\bibfield  {journal} {\bibinfo
  {journal} {Nat. Commun.}\ }\textbf {\bibinfo {volume} {12}},\ \bibinfo
  {pages} {59} (\bibinfo {year} {2021})}\BibitemShut {NoStop}%
\bibitem [{\citenamefont {Li}\ \emph {et~al.}(2018)\citenamefont {Li},
  \citenamefont {Weng}, \citenamefont {Zhang}, \citenamefont {Ding},
  \citenamefont {Zhu}, \citenamefont {Wang}, \citenamefont {Yang},
  \citenamefont {Cheng}, \citenamefont {Zhang}, \citenamefont {Li},
  \citenamefont {Lin}, \citenamefont {Chen}, \citenamefont {Han}, \citenamefont
  {Zhang}, \citenamefont {Chen}, \citenamefont {Chen}, \citenamefont {Chen},
  \citenamefont {Dong}, \citenamefont {Chen},\ and\ \citenamefont
  {Wu}}]{RN118}%
  \BibitemOpen
  \bibfield  {author} {\bibinfo {author} {\bibfnamefont {Y.}~\bibnamefont
  {Li}}, \bibinfo {author} {\bibfnamefont {Y.}~\bibnamefont {Weng}}, \bibinfo
  {author} {\bibfnamefont {J.}~\bibnamefont {Zhang}}, \bibinfo {author}
  {\bibfnamefont {J.}~\bibnamefont {Ding}}, \bibinfo {author} {\bibfnamefont
  {Y.}~\bibnamefont {Zhu}}, \bibinfo {author} {\bibfnamefont {Q.}~\bibnamefont
  {Wang}}, \bibinfo {author} {\bibfnamefont {Y.}~\bibnamefont {Yang}}, \bibinfo
  {author} {\bibfnamefont {Y.}~\bibnamefont {Cheng}}, \bibinfo {author}
  {\bibfnamefont {Q.}~\bibnamefont {Zhang}}, \bibinfo {author} {\bibfnamefont
  {P.}~\bibnamefont {Li}}, \bibinfo {author} {\bibfnamefont {J.}~\bibnamefont
  {Lin}}, \bibinfo {author} {\bibfnamefont {W.}~\bibnamefont {Chen}}, \bibinfo
  {author} {\bibfnamefont {Y.}~\bibnamefont {Han}}, \bibinfo {author}
  {\bibfnamefont {X.}~\bibnamefont {Zhang}}, \bibinfo {author} {\bibfnamefont
  {L.}~\bibnamefont {Chen}}, \bibinfo {author} {\bibfnamefont {X.}~\bibnamefont
  {Chen}}, \bibinfo {author} {\bibfnamefont {J.}~\bibnamefont {Chen}}, \bibinfo
  {author} {\bibfnamefont {S.}~\bibnamefont {Dong}}, \bibinfo {author}
  {\bibfnamefont {X.}~\bibnamefont {Chen}},\ and\ \bibinfo {author}
  {\bibfnamefont {T.}~\bibnamefont {Wu}},\ }\href
  {https://doi.org/10.1038/s41427-018-0050-5} {\bibfield  {journal} {\bibinfo
  {journal} {NPG Asia Mater.}\ }\textbf {\bibinfo {volume} {10}},\ \bibinfo
  {pages} {522} (\bibinfo {year} {2018})}\BibitemShut {NoStop}%
\bibitem [{\citenamefont {R\"odel}\ \emph {et~al.}(2014)\citenamefont
  {R\"odel}, \citenamefont {Bareille}, \citenamefont {Fortuna}, \citenamefont
  {Baumier}, \citenamefont {Bertran}, \citenamefont {Le~F\`evre}, \citenamefont
  {Gabay}, \citenamefont {Hijano~Cubelos}, \citenamefont {Rozenberg},
  \citenamefont {Maroutian}, \citenamefont {Lecoeur},\ and\ \citenamefont
  {Santander-Syro}}]{PhysRevApplied.1.051002}%
  \BibitemOpen
  \bibfield  {author} {\bibinfo {author} {\bibfnamefont {T.~C.}\ \bibnamefont
  {R\"odel}}, \bibinfo {author} {\bibfnamefont {C.}~\bibnamefont {Bareille}},
  \bibinfo {author} {\bibfnamefont {F.}~\bibnamefont {Fortuna}}, \bibinfo
  {author} {\bibfnamefont {C.}~\bibnamefont {Baumier}}, \bibinfo {author}
  {\bibfnamefont {F.}~\bibnamefont {Bertran}}, \bibinfo {author} {\bibfnamefont
  {P.}~\bibnamefont {Le~F\`evre}}, \bibinfo {author} {\bibfnamefont
  {M.}~\bibnamefont {Gabay}}, \bibinfo {author} {\bibfnamefont
  {O.}~\bibnamefont {Hijano~Cubelos}}, \bibinfo {author} {\bibfnamefont
  {M.~J.}\ \bibnamefont {Rozenberg}}, \bibinfo {author} {\bibfnamefont
  {T.}~\bibnamefont {Maroutian}}, \bibinfo {author} {\bibfnamefont
  {P.}~\bibnamefont {Lecoeur}},\ and\ \bibinfo {author} {\bibfnamefont {A.~F.}\
  \bibnamefont {Santander-Syro}},\ }\href
  {https://doi.org/10.1103/PhysRevApplied.1.051002} {\bibfield  {journal}
  {\bibinfo  {journal} {Phys. Rev. Appl.}\ }\textbf {\bibinfo {volume} {1}},\
  \bibinfo {pages} {051002(R)} (\bibinfo {year} {2014})}\BibitemShut {NoStop}%
\bibitem [{\citenamefont {He}\ \emph {et~al.}(2017)\citenamefont {He},
  \citenamefont {Jia}, \citenamefont {Hou}, \citenamefont {Wei}, \citenamefont
  {Xie}, \citenamefont {Yang}, \citenamefont {Shi}, \citenamefont {Yuan},
  \citenamefont {Shan}, \citenamefont {Zhu}, \citenamefont {Li}, \citenamefont
  {Gu}, \citenamefont {Liu}, \citenamefont {Xiang},\ and\ \citenamefont
  {Jin}}]{PhysRevB.95.054510}%
  \BibitemOpen
  \bibfield  {author} {\bibinfo {author} {\bibfnamefont {G.}~\bibnamefont
  {He}}, \bibinfo {author} {\bibfnamefont {Y.}~\bibnamefont {Jia}}, \bibinfo
  {author} {\bibfnamefont {X.}~\bibnamefont {Hou}}, \bibinfo {author}
  {\bibfnamefont {Z.}~\bibnamefont {Wei}}, \bibinfo {author} {\bibfnamefont
  {H.}~\bibnamefont {Xie}}, \bibinfo {author} {\bibfnamefont {Z.}~\bibnamefont
  {Yang}}, \bibinfo {author} {\bibfnamefont {J.}~\bibnamefont {Shi}}, \bibinfo
  {author} {\bibfnamefont {J.}~\bibnamefont {Yuan}}, \bibinfo {author}
  {\bibfnamefont {L.}~\bibnamefont {Shan}}, \bibinfo {author} {\bibfnamefont
  {B.}~\bibnamefont {Zhu}}, \bibinfo {author} {\bibfnamefont {H.}~\bibnamefont
  {Li}}, \bibinfo {author} {\bibfnamefont {L.}~\bibnamefont {Gu}}, \bibinfo
  {author} {\bibfnamefont {K.}~\bibnamefont {Liu}}, \bibinfo {author}
  {\bibfnamefont {T.}~\bibnamefont {Xiang}},\ and\ \bibinfo {author}
  {\bibfnamefont {K.}~\bibnamefont {Jin}},\ }\href
  {https://doi.org/10.1103/PhysRevB.95.054510} {\bibfield  {journal} {\bibinfo
  {journal} {Phys. Rev. B}\ }\textbf {\bibinfo {volume} {95}},\ \bibinfo
  {pages} {054510} (\bibinfo {year} {2017})}\BibitemShut {NoStop}%
\bibitem [{\citenamefont {Sun}\ \emph {et~al.}(2021)\citenamefont {Sun},
  \citenamefont {Liu}, \citenamefont {Hong}, \citenamefont {Chen},
  \citenamefont {Zhang},\ and\ \citenamefont {Xie}}]{PhysRevLett.127.086804}%
  \BibitemOpen
  \bibfield  {author} {\bibinfo {author} {\bibfnamefont {Y.}~\bibnamefont
  {Sun}}, \bibinfo {author} {\bibfnamefont {Y.}~\bibnamefont {Liu}}, \bibinfo
  {author} {\bibfnamefont {S.}~\bibnamefont {Hong}}, \bibinfo {author}
  {\bibfnamefont {Z.}~\bibnamefont {Chen}}, \bibinfo {author} {\bibfnamefont
  {M.}~\bibnamefont {Zhang}},\ and\ \bibinfo {author} {\bibfnamefont
  {Y.}~\bibnamefont {Xie}},\ }\href
  {https://doi.org/10.1103/PhysRevLett.127.086804} {\bibfield  {journal}
  {\bibinfo  {journal} {Phys. Rev. Lett.}\ }\textbf {\bibinfo {volume} {127}},\
  \bibinfo {pages} {086804} (\bibinfo {year} {2021})}\BibitemShut {NoStop}%
\bibitem [{sup()}]{supplemental}%
  \BibitemOpen
  \href@noop {} {}\bibinfo {note} {See Supplemental Material at [URL will be
  inserted by publisher] for details about the films deposited on LSAT
  substrates with different orientations, Raman spectra measurements, in-plane
  XRD measurements, and the comparison among several possible interfaces, which
  includes Refs.
  \cite{PhysRevB.101.220510,RN118,doi:10.1021/acsami.6b11791,doi:10.1143/JPSJ.71.1848}.}\BibitemShut
  {Stop}%
\bibitem [{\citenamefont {Wicklein}\ \emph {et~al.}(2012)\citenamefont
  {Wicklein}, \citenamefont {Sambri}, \citenamefont {Amoruso}, \citenamefont
  {Wang}, \citenamefont {Bruzzese}, \citenamefont {Koehl},\ and\ \citenamefont
  {Dittmann}}]{doi:10.1063/1.4754112}%
  \BibitemOpen
  \bibfield  {author} {\bibinfo {author} {\bibfnamefont {S.}~\bibnamefont
  {Wicklein}}, \bibinfo {author} {\bibfnamefont {A.}~\bibnamefont {Sambri}},
  \bibinfo {author} {\bibfnamefont {S.}~\bibnamefont {Amoruso}}, \bibinfo
  {author} {\bibfnamefont {X.}~\bibnamefont {Wang}}, \bibinfo {author}
  {\bibfnamefont {R.}~\bibnamefont {Bruzzese}}, \bibinfo {author}
  {\bibfnamefont {A.}~\bibnamefont {Koehl}},\ and\ \bibinfo {author}
  {\bibfnamefont {R.}~\bibnamefont {Dittmann}},\ }\href
  {https://doi.org/10.1063/1.4754112} {\bibfield  {journal} {\bibinfo
  {journal} {Appl. Phys. Lett.}\ }\textbf {\bibinfo {volume} {101}},\ \bibinfo
  {pages} {131601} (\bibinfo {year} {2012})}\BibitemShut {NoStop}%
\bibitem [{\citenamefont {Breckenfeld}\ \emph {et~al.}(2013)\citenamefont
  {Breckenfeld}, \citenamefont {Bronn}, \citenamefont {Karthik}, \citenamefont
  {Damodaran}, \citenamefont {Lee}, \citenamefont {Mason},\ and\ \citenamefont
  {Martin}}]{PhysRevLett.110.196804}%
  \BibitemOpen
  \bibfield  {author} {\bibinfo {author} {\bibfnamefont {E.}~\bibnamefont
  {Breckenfeld}}, \bibinfo {author} {\bibfnamefont {N.}~\bibnamefont {Bronn}},
  \bibinfo {author} {\bibfnamefont {J.}~\bibnamefont {Karthik}}, \bibinfo
  {author} {\bibfnamefont {A.~R.}\ \bibnamefont {Damodaran}}, \bibinfo {author}
  {\bibfnamefont {S.}~\bibnamefont {Lee}}, \bibinfo {author} {\bibfnamefont
  {N.}~\bibnamefont {Mason}},\ and\ \bibinfo {author} {\bibfnamefont {L.~W.}\
  \bibnamefont {Martin}},\ }\href
  {https://doi.org/10.1103/PhysRevLett.110.196804} {\bibfield  {journal}
  {\bibinfo  {journal} {Phys. Rev. Lett.}\ }\textbf {\bibinfo {volume} {110}},\
  \bibinfo {pages} {196804} (\bibinfo {year} {2013})}\BibitemShut {NoStop}%
\bibitem [{\citenamefont {Ojeda-G-P}\ \emph {et~al.}(2018)\citenamefont
  {Ojeda-G-P}, \citenamefont {D\"{o}beli},\ and\ \citenamefont
  {Lippert}}]{10.1002/admi.201701062}%
  \BibitemOpen
  \bibfield  {author} {\bibinfo {author} {\bibfnamefont {A.}~\bibnamefont
  {Ojeda-G-P}}, \bibinfo {author} {\bibfnamefont {M.}~\bibnamefont
  {D\"{o}beli}},\ and\ \bibinfo {author} {\bibfnamefont {T.}~\bibnamefont
  {Lippert}},\ }\href {https://doi.org/10.1002/admi.201701062} {\bibfield
  {journal} {\bibinfo  {journal} {Adv. Mater. Interfaces}\ }\textbf {\bibinfo
  {volume} {5}},\ \bibinfo {pages} {1701062} (\bibinfo {year}
  {2018})}\BibitemShut {NoStop}%
\bibitem [{\citenamefont {Canulescu}\ \emph {et~al.}(2017)\citenamefont
  {Canulescu}, \citenamefont {D\"obeli}, \citenamefont {Yao}, \citenamefont
  {Lippert}, \citenamefont {Amoruso},\ and\ \citenamefont
  {Schou}}]{PhysRevMaterials.1.073402}%
  \BibitemOpen
  \bibfield  {author} {\bibinfo {author} {\bibfnamefont {S.}~\bibnamefont
  {Canulescu}}, \bibinfo {author} {\bibfnamefont {M.}~\bibnamefont {D\"obeli}},
  \bibinfo {author} {\bibfnamefont {X.}~\bibnamefont {Yao}}, \bibinfo {author}
  {\bibfnamefont {T.}~\bibnamefont {Lippert}}, \bibinfo {author} {\bibfnamefont
  {S.}~\bibnamefont {Amoruso}},\ and\ \bibinfo {author} {\bibfnamefont
  {J.}~\bibnamefont {Schou}},\ }\href
  {https://doi.org/10.1103/PhysRevMaterials.1.073402} {\bibfield  {journal}
  {\bibinfo  {journal} {Phys. Rev. Materials}\ }\textbf {\bibinfo {volume}
  {1}},\ \bibinfo {pages} {073402} (\bibinfo {year} {2017})}\BibitemShut
  {NoStop}%
\bibitem [{\citenamefont {Schou}(2009)}]{SCHOU20095191}%
  \BibitemOpen
  \bibfield  {author} {\bibinfo {author} {\bibfnamefont {J.}~\bibnamefont
  {Schou}},\ }\href {https://doi.org/10.1016/j.apsusc.2008.10.101} {\bibfield
  {journal} {\bibinfo  {journal} {Appl. Surf. Sci.}\ }\textbf {\bibinfo
  {volume} {255}},\ \bibinfo {pages} {5191} (\bibinfo {year}
  {2009})}\BibitemShut {NoStop}%
\bibitem [{\citenamefont {Packwood}\ \emph {et~al.}(2013)\citenamefont
  {Packwood}, \citenamefont {Shiraki},\ and\ \citenamefont
  {Hitosugi}}]{PhysRevLett.111.036101}%
  \BibitemOpen
  \bibfield  {author} {\bibinfo {author} {\bibfnamefont {D.~M.}\ \bibnamefont
  {Packwood}}, \bibinfo {author} {\bibfnamefont {S.}~\bibnamefont {Shiraki}},\
  and\ \bibinfo {author} {\bibfnamefont {T.}~\bibnamefont {Hitosugi}},\ }\href
  {https://doi.org/10.1103/PhysRevLett.111.036101} {\bibfield  {journal}
  {\bibinfo  {journal} {Phys. Rev. Lett.}\ }\textbf {\bibinfo {volume} {111}},\
  \bibinfo {pages} {036101} (\bibinfo {year} {2013})}\BibitemShut {NoStop}%
\bibitem [{\citenamefont {Schneider}\ \emph {et~al.}(2010)\citenamefont
  {Schneider}, \citenamefont {Esposito}, \citenamefont {Marozau}, \citenamefont
  {Conder}, \citenamefont {Doebeli}, \citenamefont {Hu}, \citenamefont
  {Mallepell}, \citenamefont {Wokaun},\ and\ \citenamefont
  {Lippert}}]{doi:10.1063/1.3515849}%
  \BibitemOpen
  \bibfield  {author} {\bibinfo {author} {\bibfnamefont {C.~W.}\ \bibnamefont
  {Schneider}}, \bibinfo {author} {\bibfnamefont {M.}~\bibnamefont {Esposito}},
  \bibinfo {author} {\bibfnamefont {I.}~\bibnamefont {Marozau}}, \bibinfo
  {author} {\bibfnamefont {K.}~\bibnamefont {Conder}}, \bibinfo {author}
  {\bibfnamefont {M.}~\bibnamefont {Doebeli}}, \bibinfo {author} {\bibfnamefont
  {Y.}~\bibnamefont {Hu}}, \bibinfo {author} {\bibfnamefont {M.}~\bibnamefont
  {Mallepell}}, \bibinfo {author} {\bibfnamefont {A.}~\bibnamefont {Wokaun}},\
  and\ \bibinfo {author} {\bibfnamefont {T.}~\bibnamefont {Lippert}},\ }\href
  {https://doi.org/10.1063/1.3515849} {\bibfield  {journal} {\bibinfo
  {journal} {Appl. Phys. Lett.}\ }\textbf {\bibinfo {volume} {97}},\ \bibinfo
  {pages} {192107} (\bibinfo {year} {2010})}\BibitemShut {NoStop}%
\bibitem [{\citenamefont {Hilti}(1968)}]{RN120}%
  \BibitemOpen
  \bibfield  {author} {\bibinfo {author} {\bibfnamefont {E.}~\bibnamefont
  {Hilti}},\ }\href {https://doi.org/10.1007/BF00624245} {\bibfield  {journal}
  {\bibinfo  {journal} {Naturwissenschaften}\ }\textbf {\bibinfo {volume}
  {55}},\ \bibinfo {pages} {130} (\bibinfo {year} {1968})}\BibitemShut
  {NoStop}%
\bibitem [{\citenamefont {Valeeva}\ and\ \citenamefont
  {Kostenko}(2017)}]{Valeeva2017NPCM}%
  \BibitemOpen
  \bibfield  {author} {\bibinfo {author} {\bibfnamefont {A.~A.}\ \bibnamefont
  {Valeeva}}\ and\ \bibinfo {author} {\bibfnamefont {M.~G.}\ \bibnamefont
  {Kostenko}},\ }\href {https://doi.org/10.17586/2220-8054-2017-8-6-816-822}
  {\bibfield  {journal} {\bibinfo  {journal} {Nanosyst.: Phys., Chem., Math.}\
  }\textbf {\bibinfo {volume} {8}},\ \bibinfo {pages} {816} (\bibinfo {year}
  {2017})}\BibitemShut {NoStop}%
\bibitem [{\citenamefont {Valeeva}\ \emph {et~al.}(2018)\citenamefont
  {Valeeva}, \citenamefont {Kostenko}, \citenamefont {Nazarova}, \citenamefont
  {Gerasimov},\ and\ \citenamefont {Rempel}}]{RN119}%
  \BibitemOpen
  \bibfield  {author} {\bibinfo {author} {\bibfnamefont {A.~A.}\ \bibnamefont
  {Valeeva}}, \bibinfo {author} {\bibfnamefont {M.~G.}\ \bibnamefont
  {Kostenko}}, \bibinfo {author} {\bibfnamefont {S.~Z.}\ \bibnamefont
  {Nazarova}}, \bibinfo {author} {\bibfnamefont {E.~Y.}\ \bibnamefont
  {Gerasimov}},\ and\ \bibinfo {author} {\bibfnamefont {A.~A.}\ \bibnamefont
  {Rempel}},\ }\href {https://doi.org/10.1134/S0020168518060158} {\bibfield
  {journal} {\bibinfo  {journal} {Inorg. Mater.}\ }\textbf {\bibinfo {volume}
  {54}},\ \bibinfo {pages} {568} (\bibinfo {year} {2018})}\BibitemShut
  {NoStop}%
\bibitem [{\citenamefont {Inaba}\ \emph {et~al.}(2013)\citenamefont {Inaba},
  \citenamefont {Kobayashi}, \citenamefont {Uehara}, \citenamefont {Okada},
  \citenamefont {Reddy},\ and\ \citenamefont {Endo}}]{Inaba2013AMPC}%
  \BibitemOpen
  \bibfield  {author} {\bibinfo {author} {\bibfnamefont {K.}~\bibnamefont
  {Inaba}}, \bibinfo {author} {\bibfnamefont {S.}~\bibnamefont {Kobayashi}},
  \bibinfo {author} {\bibfnamefont {K.}~\bibnamefont {Uehara}}, \bibinfo
  {author} {\bibfnamefont {A.}~\bibnamefont {Okada}}, \bibinfo {author}
  {\bibfnamefont {S.~L.}\ \bibnamefont {Reddy}},\ and\ \bibinfo {author}
  {\bibfnamefont {T.}~\bibnamefont {Endo}},\ }\href
  {https://doi.org/10.4236/ampc.2013.31A010} {\bibfield  {journal} {\bibinfo
  {journal} {Adv. Mater. Phys. Chem.}\ }\textbf {\bibinfo {volume} {3}},\
  \bibinfo {pages} {72} (\bibinfo {year} {2013})}\BibitemShut {NoStop}%
\bibitem [{\citenamefont {Ding}\ \emph {et~al.}(2016)\citenamefont {Ding},
  \citenamefont {Dwaraknath}, \citenamefont {Garten}, \citenamefont {Ndione},
  \citenamefont {Ginley},\ and\ \citenamefont
  {Persson}}]{doi:10.1021/acsami.6b01630}%
  \BibitemOpen
  \bibfield  {author} {\bibinfo {author} {\bibfnamefont {H.}~\bibnamefont
  {Ding}}, \bibinfo {author} {\bibfnamefont {S.~S.}\ \bibnamefont
  {Dwaraknath}}, \bibinfo {author} {\bibfnamefont {L.}~\bibnamefont {Garten}},
  \bibinfo {author} {\bibfnamefont {P.}~\bibnamefont {Ndione}}, \bibinfo
  {author} {\bibfnamefont {D.}~\bibnamefont {Ginley}},\ and\ \bibinfo {author}
  {\bibfnamefont {K.~A.}\ \bibnamefont {Persson}},\ }\href
  {https://doi.org/10.1021/acsami.6b01630} {\bibfield  {journal} {\bibinfo
  {journal} {ACS Appl. Mater. Interfaces}\ }\textbf {\bibinfo {volume} {8}},\
  \bibinfo {pages} {13086} (\bibinfo {year} {2016})}\BibitemShut {NoStop}%
\bibitem [{\citenamefont {Werthamer}\ \emph {et~al.}(1966)\citenamefont
  {Werthamer}, \citenamefont {Helfand},\ and\ \citenamefont
  {Hohenberg}}]{PhysRev.147.295}%
  \BibitemOpen
  \bibfield  {author} {\bibinfo {author} {\bibfnamefont {N.~R.}\ \bibnamefont
  {Werthamer}}, \bibinfo {author} {\bibfnamefont {E.}~\bibnamefont {Helfand}},\
  and\ \bibinfo {author} {\bibfnamefont {P.~C.}\ \bibnamefont {Hohenberg}},\
  }\href {https://doi.org/10.1103/PhysRev.147.295} {\bibfield  {journal}
  {\bibinfo  {journal} {Phys. Rev.}\ }\textbf {\bibinfo {volume} {147}},\
  \bibinfo {pages} {295} (\bibinfo {year} {1966})}\BibitemShut {NoStop}%
\bibitem [{\citenamefont {Wei}\ \emph {et~al.}(2019)\citenamefont {Wei},
  \citenamefont {He}, \citenamefont {Hu}, \citenamefont {Feng}, \citenamefont
  {Wei}, \citenamefont {Ho}, \citenamefont {Li}, \citenamefont {Yuan},
  \citenamefont {Xi}, \citenamefont {Wang}, \citenamefont {Chen}, \citenamefont
  {Zhu}, \citenamefont {Zhou}, \citenamefont {Dong}, \citenamefont {Pi},
  \citenamefont {Kusmartseva}, \citenamefont {Kusmartsev}, \citenamefont
  {Zhao},\ and\ \citenamefont {Jin}}]{PhysRevB.100.184509}%
  \BibitemOpen
  \bibfield  {author} {\bibinfo {author} {\bibfnamefont {Z.}~\bibnamefont
  {Wei}}, \bibinfo {author} {\bibfnamefont {G.}~\bibnamefont {He}}, \bibinfo
  {author} {\bibfnamefont {W.}~\bibnamefont {Hu}}, \bibinfo {author}
  {\bibfnamefont {Z.}~\bibnamefont {Feng}}, \bibinfo {author} {\bibfnamefont
  {X.}~\bibnamefont {Wei}}, \bibinfo {author} {\bibfnamefont {C.~Y.}\
  \bibnamefont {Ho}}, \bibinfo {author} {\bibfnamefont {Q.}~\bibnamefont {Li}},
  \bibinfo {author} {\bibfnamefont {J.}~\bibnamefont {Yuan}}, \bibinfo {author}
  {\bibfnamefont {C.}~\bibnamefont {Xi}}, \bibinfo {author} {\bibfnamefont
  {Z.}~\bibnamefont {Wang}}, \bibinfo {author} {\bibfnamefont {Q.}~\bibnamefont
  {Chen}}, \bibinfo {author} {\bibfnamefont {B.}~\bibnamefont {Zhu}}, \bibinfo
  {author} {\bibfnamefont {F.}~\bibnamefont {Zhou}}, \bibinfo {author}
  {\bibfnamefont {X.}~\bibnamefont {Dong}}, \bibinfo {author} {\bibfnamefont
  {L.}~\bibnamefont {Pi}}, \bibinfo {author} {\bibfnamefont {A.}~\bibnamefont
  {Kusmartseva}}, \bibinfo {author} {\bibfnamefont {F.~V.}\ \bibnamefont
  {Kusmartsev}}, \bibinfo {author} {\bibfnamefont {Z.}~\bibnamefont {Zhao}},\
  and\ \bibinfo {author} {\bibfnamefont {K.}~\bibnamefont {Jin}},\ }\href
  {https://doi.org/10.1103/PhysRevB.100.184509} {\bibfield  {journal} {\bibinfo
   {journal} {Phys. Rev. B}\ }\textbf {\bibinfo {volume} {100}},\ \bibinfo
  {pages} {184509} (\bibinfo {year} {2019})}\BibitemShut {NoStop}%
\bibitem [{\citenamefont {Hu}(2016)}]{HU2016561}%
  \BibitemOpen
  \bibfield  {author} {\bibinfo {author} {\bibfnamefont {J.}~\bibnamefont
  {Hu}},\ }\href {https://doi.org/10.1007/s11434-016-1037-7} {\bibfield
  {journal} {\bibinfo  {journal} {Sci. Bull.}\ }\textbf {\bibinfo {volume}
  {61}},\ \bibinfo {pages} {561} (\bibinfo {year} {2016})}\BibitemShut
  {NoStop}%
\bibitem [{\citenamefont {Xu}\ \emph {et~al.}(2017)\citenamefont {Xu},
  \citenamefont {Salvador},\ and\ \citenamefont
  {Kitchin}}]{doi:10.1021/acsami.6b11791}%
  \BibitemOpen
  \bibfield  {author} {\bibinfo {author} {\bibfnamefont {Z.}~\bibnamefont
  {Xu}}, \bibinfo {author} {\bibfnamefont {P.}~\bibnamefont {Salvador}},\ and\
  \bibinfo {author} {\bibfnamefont {J.~R.}\ \bibnamefont {Kitchin}},\ }\href
  {https://doi.org/10.1021/acsami.6b11791} {\bibfield  {journal} {\bibinfo
  {journal} {ACS Appl. Mater. Interfaces}\ }\textbf {\bibinfo {volume} {9}},\
  \bibinfo {pages} {4106} (\bibinfo {year} {2017})}\BibitemShut {NoStop}%
\bibitem [{\citenamefont {Isobe}\ and\ \citenamefont
  {Ueda}(2002)}]{doi:10.1143/JPSJ.71.1848}%
  \BibitemOpen
  \bibfield  {author} {\bibinfo {author} {\bibfnamefont {M.}~\bibnamefont
  {Isobe}}\ and\ \bibinfo {author} {\bibfnamefont {Y.}~\bibnamefont {Ueda}},\
  }\href {https://doi.org/10.1143/JPSJ.71.1848} {\bibfield  {journal} {\bibinfo
   {journal} {J. Phys. Soc. Jpn.}\ }\textbf {\bibinfo {volume} {71}},\ \bibinfo
  {pages} {1848} (\bibinfo {year} {2002})}\BibitemShut {NoStop}%
\end{thebibliography}%

\end{document}